\documentclass{article}
\usepackage{authblk}
\usepackage[utf8]{inputenc}
\usepackage[round]{natbib}
\usepackage[pdftex]{graphicx}
\usepackage{amsmath,amsthm,amsfonts}
\usepackage{diagbox}
\usepackage{lscape}   
\usepackage{rotating}
\usepackage{rotfloat}
\usepackage{amsthm}
\usepackage{multirow}
\usepackage{booktabs}
\usepackage{ctable}
\usepackage{threeparttable}
\usepackage{booktabs}
\usepackage{framed}
\usepackage{color}
\usepackage[colorlinks=true,linkcolor=blue,allcolors=blue]{hyperref}

\usepackage{soul}

\newtheorem{proposition}{Proposition}
\title{Wavelet shrinkage based on the raised cosine prior}

\author{Juliana Marchesi Reina}
\author{Alex Rodrigo dos S. Sousa}

\affil{Universidade Estadual de Campinas (UNICAMP)\\ Departament of Statistics, Brazil \thanks{Reina (j271079@dac.unicamp.br) and Sousa (asousa@unicamp.br)}}

\date{\today} 

\begin{document}

\maketitle

\vspace{-1.0cm}
\begin{abstract}
We propose a Bayesian shrinkage rule to estimate the wavelet coefficients in a nonparametric regression model with Gaussian errors, based on a mixture of a point mass function at zero and a symmetric, zero-centered raised cosine distribution prior. The proposed rule outperformed established shrinkage and thresholding methods in specific scenarios of signal-to-noise ratio and sample size values in conducted simulation studies involving the so-called Donoho and Johnstone test functions. Statistical properties of the rule, such as squared bias, variance, and risks, are analyzed, and two illustrations in real datasets are 
provided. 
  \\

  \end{abstract}

  

\section{Introduction}
In the standard nonparametric regression problem, whose goal is to estimate an unknown function from observed data with no assumption about its shape, a conventional strategy consists of representing it through linear combinations of basis functions of the assumed function space to which the function belongs. In this way, the problem of estimating the function is equivalent to estimating the coefficients of the expansion.

Although there are several basis to consider, such as polynomials, splines (and their variations), and the Fourier basis, the wavelet basis has been widely employed in several research areas, with applications ranging from image processing and digital signal analysis to time series analysis. It is particularly attractive in nonparametric regression, mainly due to a key characteristic: wavelets are well localized in both the time and scale domains. As a consequence, the coefficients of the expansion of the target function tend to be sparse, concentrating most of the relevant information in a few large-magnitude coefficients. This not only facilitates the representation of the function but also enhances the estimation process of the associated coefficients. See \cite{daubechies1992ten} and \cite{stephane-mallat-2009} for a theoretical overview of wavelets, but also \cite{vidakovic-1999} and \cite{nason-2008} for applications of wavelets in statistics.

The sparsity property has motivated the development of several strategies for estimating wavelet coefficients in signals contaminated by noise. The general idea is to reduce the magnitude of the empirical wavelet coefficients obtained by applying the discrete wavelet transform (DWT) to the data, thereby estimating the wavelet coefficients. Thresholding techniques aim to establish a threshold within which empirical coefficients are set to zero due to their low magnitude. Outside this threshold, there are different strategies for handling the significant coefficients, which may either preserve their magnitude or shrink them. See \cite{DJ-1994,DJ-1995} for seminal works on wavelet shrinkage. Bayesian approaches have also been developed to incorporate prior information associated with the wavelet coefficients, such as sparsity and its support, by defining a prior distribution for these coefficients. Several priors for the wavelet coefficients were available in the literature, most of which acted as spike and slab distributions. \cite{chipman-1997} proposed a mixture of normal densities, but the usual structure is to consider a mixture of a point mass function at zero and a symmetric and unimodal distribution, such as the double exponential, uniform, double Weibull, beta and logistic distributions, proposed by \cite{vidakovic-ruggeri-bams}, \cite{angelini-vidakovic-2004}, \cite{remenyi-2015}, \cite{alex-beta} and \cite{alex-logistica} respectively. See also the recent work of \cite{vimalajeewa-2023} for a Bayesian shrinkage rule for data with a high noise level.

In this sense, we propose a particular prior for the wavelet coefficients, composed of a mixture of a point mass function at zero and a symmetric, raised cosine distribution centered around zero. Its hyperparameters allow the control of the degree of shrinkage imposed on the empirical coefficients, and the associated shrinkage rule surpassed the standard methods in some scenarios of the simulation studies in terms of averaged mean squared and averaged median absolute errors, mainly for very noisy data. Moreover, the shrinkage rule under a raised cosine prior can be viewed as the limit of the rule proposed by \cite{angelini-vidakovic-2004}, based on a mixture of a point mass function at zero and the uniform distribution, which is minimax for the class of symmetric and unimodal priors for the wavelet coefficients.   

This paper is organized as follows: Section 2 defines the nonparametric regression model in the time and wavelet domains, and proposes a prior distribution for the wavelet coefficients based on the raised cosine distribution. The associated Bayesian shrinkage rule and its statistical properties, such as squared bias, variance, and risks, are developed in Section 3. Simulation studies to evaluate the performance of the proposed rule and to compare it with standard methods are analyzed in Section 4. Section 5 presents two real data illustrations, one involving a temperature anomaly time series and the other an electroencephalogram (EEG) dataset. Final considerations and further remarks are given in Section 6.

\section{Statistical model}
Assume a sample of $n = 2^{J}$ observations $(x_1,y_1), \dots, (x_n,y_n)$, with $J \in \mathbb{N}$ and consider the following nonparametric regression model
\begin{equation} \label{time_model} 
    y_i = f(x_i) + \varepsilon_i, \quad i = 1, \dots, n, 
\end{equation}

\noindent where $f \in \mathbb{L}_2(\mathbb{R}) = \{f:\int f^2(x) < \infty\}$ is an unknown function and $\varepsilon_i$ are independent and identically distributed (IID) random variables with a normal distribution with zero mean and unknown common variance $\sigma^2$, with $\sigma > 0$. 
The goal is to estimate the function $f$ without making any assumptions about its functional structure. The standard nonparametric procedure is to expand $f$ in basis functions, such as polynomials, splines, and their variants, the Fourier basis, wavelets, and others, and estimate the coefficients of the linear combination. See \cite{takezawa-2005} for more details about nonparametric regression.

We consider in this work the expansion of $f$ in a wavelet basis,
\begin{equation} \label{expan}
f(x) = \sum_{j,k \in \mathbb{Z}}\theta_{j,k} \psi_{j,k}(x), 
\end{equation}
where $\{\psi_{j,k}(x) = 2^{j/2} \psi(2^j x - k),j,k \in \mathbb{Z} \}$ is an orthonormal wavelet basis for $\mathbb{L}_2(\mathbb{R})$ constructed by dilations $j$ and translates $k$ of a function $\psi$ called a wavelet or mother wavelet and $\theta_{j,k} \in \mathbb{R}$ are wavelet coefficients that describe features of $f$ at spatial locations $2^{-j}k$ and scales $2^j$ or resolution levels $j$. Thus, according to the representation \eqref{expan}, the problem of estimating the function $f$ is reduced to the problem of estimating the wavelet coefficients $\theta_{j,k}$. For simplicity, the subscripts $ j$ and $ k$ will be dropped in the text without loss of interpretation. The representation of $f$ in a wavelet basis is interesting due to the sparsity characteristic of the wavelet coefficients in \eqref{expan}, i.e., the main features of $f$ can be stored in a few nonzero coefficients. 
 
In order to estimate the wavelet coefficients, we apply a DWT on the observations to transform them to the wavelet domain. First, we write the model \eqref{time_model} in vector notation as
\begin{equation} \label{vec_model}
    \boldsymbol{y} = \boldsymbol{f} + \boldsymbol{\varepsilon},
\end{equation}
\noindent where $\boldsymbol{y} = (y_1, \dots, y_n)^{\text{T}}$, $\boldsymbol{f} = (f(x_1), \dots, f(x_n))^{\text{T}}$, and $\boldsymbol{\varepsilon} = (\varepsilon_1, \dots, \varepsilon_n)^{\text{T}}$. The application of the DWT can be represented by an orthogonal transformation matrix $\boldsymbol{W}$ with dimension $n \times n$, which is applied on both sides of \eqref{vec_model}, obtaining the additive model in wavelet domain
\begin{equation}\label{wav_model}
    \boldsymbol{d} = \boldsymbol{\theta} + \boldsymbol{\epsilon},
\end{equation}
where $\boldsymbol{d} = \boldsymbol{W}\boldsymbol{y}= (d_1, \ldots, d_n)^{\text{T}}$ is the vector with the empirical (observed) wavelet coefficients, $\boldsymbol{\theta} = \boldsymbol{W}\boldsymbol{f} = (\theta_1, \ldots, \theta_n)^{\text{T}}$ is the vector with the unknown wavelet coefficients and $\boldsymbol{\epsilon} = \boldsymbol{W}\boldsymbol{\varepsilon} = (\epsilon_1, \ldots, \epsilon _n)^{\text{T}}$ is the vector with the random errors, which remain IID normally distributed with zero mean and common variance $\sigma^2$. Although we described the DWT procedure in terms of matrix multiplication for pedagogical purposes, fast algorithms are typically used to perform it, such as the pyramidal algorithm (see \cite{nason-2008}). 

The wavelet coefficients with larger magnitudes are associated with important regions of the function to be estimated, such as discontinuities, local maxima and minima, or other abrupt variations in the structure of $f$. On the other hand, the coefficients with smaller magnitudes represent the smoother and less detailed parts of the function. In this sense, the sparse characteristic of $\boldsymbol{\theta}$ not only simplifies the problem of estimating $f$ but also enables us to use a shrinkage or thresholding rule $\delta(d)$, which are particularly effective in separating signal and noise at different scales of the wavelet decomposition by reducing the magnitudes of the empirical coefficients $d$ in order to estimate $\theta$, i.e
$$\hat{\theta} = \delta(d),$$
where $\hat{\theta}$ is the estimate of $\theta$.

The main difference between classical and Bayesian methods in wavelet shrinkage lies in how the inherent sparsity of these coefficients is handled. In classical methods, the coefficients are shrunk to zero based on predefined thresholding rules, exploiting sparsity but without explicitly modeling the probabilistic behavior of the wavelet coefficients. In contrast, the Bayesian approach explicitly assumes that sparsity is prior knowledge, incorporating it into the model through a prior distribution designed to reflect this characteristic. Thus, the challenge in Bayesian methods is to define an appropriate prior that balances concentration around zero (sparsity) and the probability that some coefficients are significantly different from zero. 

A commonly used approach for this purpose consists in assigning a prior distribution $\pi(\cdot)$ to each wavelet coefficient, $\theta$, defined as a mixture of a point mass function at zero $\delta(\cdot)$ and a unimodal and symmetric around zero density function $g(\cdot)$,
\begin{equation} \label{eq:prior}
    \pi(\theta; \alpha) = \alpha \delta_0 (\theta) + (1-\alpha) g(\theta),
\end{equation}
where $\alpha \in (0,1)$ is a hyperparameter. In this work, we propose the density function $g(\theta)$ to be the symmetric around zero raised cosine density,
\begin{equation} \label{RC}
    g(\theta; \tau) = \frac{1}{2 \tau} \left[ 1 + \cos \left( \frac{\pi \theta}{\tau}\right) \right] \mathbb{I}_{(\tau, ~ \tau)} (\theta),
\end{equation}
\noindent where  $\tau > 0$ is the scale hyperparameter and $\mathbb{I}_{(\tau, ~ \tau)} (.)$ is the indicator function over the interval $(\tau, ~ \tau)$. This distribution, also referred to in the literature by the Bickel distribution due to its development in \citet{bickel}, was originally introduced in the context of minimax estimation of the mean of a normal distribution in the case where the mean is bounded on $(- \tau, ~ \tau)$. In this setting, the author demonstrated that as $\tau \rightarrow{\infty}$, this distribution approaches the least favorable prior within the class of symmetric and unimodal distributions on $(-\tau, \tau)$, denoted by $\Gamma_{SU}[\tau, ~ \tau]$. Two decades later, \cite{angelini-vidakovic-2004} developed the $\Gamma$-minimax wavelet shrinkage rule, where $\Gamma = \{\pi(\theta;\alpha) = \alpha\delta_0(\theta) + (1-\alpha)g(\theta), g(\theta) \in \Gamma_{SU}[-\tau,\tau]\}$, i.e, they proposed a minimax shrinkage rule in the class of priors of the form \eqref{eq:prior} but $g(\theta)$ being any symmetric and unimodal density function on $(-\tau,\tau)$. The $\Gamma$-minimax shrinkage rule is associated with the prior \eqref{eq:prior} and $g(\theta)$ being the uniform density function on $(-\tau, \tau)$, which is called the least favorable prior. Furthermore, the authors note that when $\tau$ is large, the least favorable prior is close to the prior \eqref{eq:prior} with $g(\theta)$ given by \eqref{RC}.  Thus, in the context of wavelet coefficient estimation, assuming the proposed prior \eqref{eq:prior} and \eqref{RC}, the associated Bayesian rule approximates the $\Gamma$-minimax shrinkage rule. 

\begin{figure}
    \centering
    \includegraphics[width=1\linewidth]{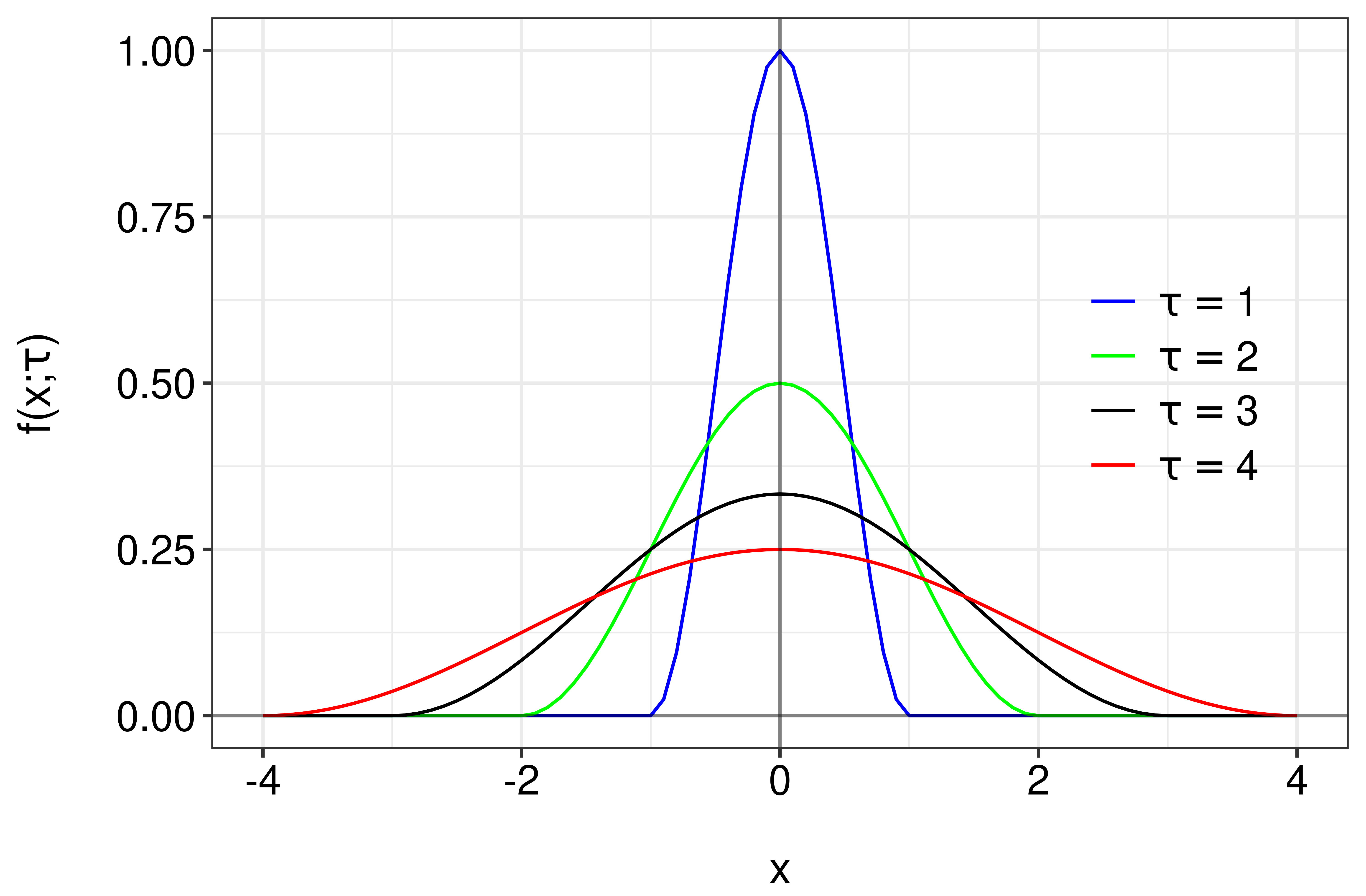} 
    \caption{Raised cosine density function for $\tau \in \{1,2,3,4\}$.}
    \label{rcdensity:app}
\end{figure}

Figure \ref{rcdensity:app} shows the raised cosine density function \eqref{RC} for $\tau \in \{1,2,3,4\}$. Note that the parameter $\tau$, in addition to defining the scale of the distribution, also determines the interval in which the density is nonzero, i.e., the support of the distribution. This unique characteristic will later play a crucial role in defining the estimation method for this parameter. Furthermore, as $\tau$ increases, the cosine distribution becomes more dispersed, characterized by greater amplitude and lower kurtosis, which increases the variance and reduces the density concentration around the mean. On the other hand, smaller values of $\tau$ result in a more concentrated distribution, with a sharp peak precisely at the mean, indicating lower variance and higher kurtosis. 

Moreover, the control that the parameter $\tau$ exerts over the support and scale of the distribution directly influences the degree of shrinkage applied to the empirical wavelet coefficients and the sparsity of the estimated coefficient vector. A higher value of $\tau$ in the raised cosine distribution, used as a prior, implies a lower probability concentration around zero, resulting in more dispersed coefficients and, consequently, less shrinkage.

\section{Shrinkage rule and its properties}
To derive the shrinkage rule $\delta(\cdot)$ associated to the model \eqref{wav_model} and prior distribution \eqref{eq:prior} and \eqref{RC} that define the Bayesian estimator of the wavelet coefficient $\theta$, we assume the quadratic loss function $L(\delta, \theta) = (\delta - \theta)^2$, whose Bayes risk is minimized by taking the posterior mean of $\theta$, i.e, $\delta(d) = \mathbb{E}_{\pi} (\theta | d)$, see \cite{berger-1985}. Proposition \ref{prop} presents an expression of the proposed shrinkage rule in terms of the cumulative and probability density functions of the standard normal distribution. The proof is in the Appendix.

\begin{proposition} \label{prop}
    Consider the model \eqref{wav_model} and let $\pi(\theta; \alpha, \tau) = \alpha \delta_0 (\theta) + (1-\alpha) g(\theta; \tau)$ be the prior distribution for $\theta$, where $g(\theta; \tau)$ is the raised cosine probability density function \eqref{RC}. Then, the Bayesian shrinkage rule $\delta(d)$ under the quadratic loss function is

  \begin{equation} \label{eq:rule}
      \delta(d) = \frac{(1-\alpha) \frac{\sigma}{2 \tau} \left [\phi \left( \frac{- \tau - d}{\sigma} \right) - \phi \left( \frac{\tau - d}{\sigma} \right) \right] + \frac{d}{2 \tau} \left [\Phi \left( \frac{\tau - d}{\sigma} \right) - \Phi \left( \frac{- \tau - d}{\sigma} \right) \right] + I_1} {\alpha \frac{1}{\sigma} \phi(\frac{d}{\sigma}) + (1-\alpha) \frac{1}{2 \tau} \left [\Phi \left( \frac{\tau - d}{\sigma} \right) - \Phi \left( \frac{- \tau - d}{\sigma} \right) \right] + I_2}, 
  \end{equation}

\noindent where $\phi(.)$ and $\Phi(.)$ denote the probability density function and the cumulative distribution function of the standard normal distribution, respectively, and
  \begin{equation*}
      \begin{aligned}
          I_1 & = \int_{\frac{- \tau - d}{\sigma}}^{\frac{\tau - d}{\sigma}} (\sigma u + d) 
          \frac{1}{2 \tau} \cos\left(\frac{\pi (\sigma u + d)}{\tau}\right) \phi(u) \, du, \\
          I_2 & = \int_{\frac{- \tau - d}{\sigma}}^{\frac{\tau - d}{\sigma}} \frac{1}{2 \tau} 
          \cos\left(\frac{\pi (\sigma u + d)}{\tau}\right) \phi(u) \, du.
      \end{aligned}
  \end{equation*}
\end{proposition}

\begin{figure}
    \centering
    \includegraphics[width=0.8\linewidth]{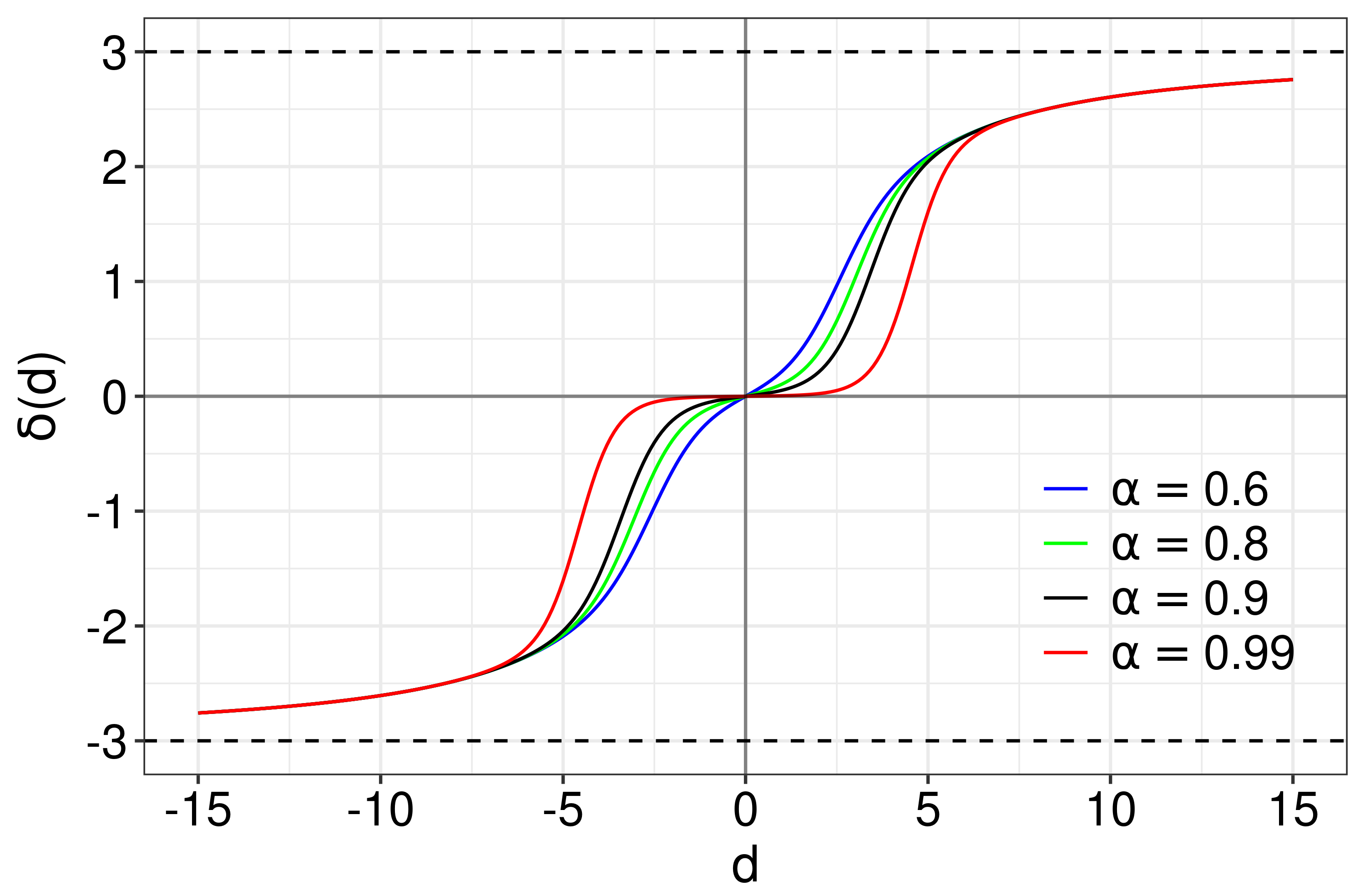} 
    \caption{Shrinkage rules under raised cosine prior for $\sigma = 1$, $\tau = 3$ and $\alpha \in \{0.6; 0.8; 0.9; 0.99\}$.}
    \label{rules1:app}
\end{figure}

The integrals $I_1$ and $I_2$ of \eqref{eq:rule} are obtained numerically. Figure \ref{rules1:app} presents the shrinkage rules obtained for cases where $\tau = 3$ and $\alpha \in \{0.6; 0.8; 0.9; 0.99 \}$. In both cases, we considered $\sigma = 1$. It is observed that, as $\alpha$ increases, the range of $d$ values in which the coefficients are shrunk to zero also increases. This is reflected in the behavior of the curves, which exhibit a broader transition around $d=0$, indicating that higher values of $\alpha$ are associated with a greater degree of shrinkage applied to wavelet coefficients of smaller amplitude, which is reasonable since $\alpha$ is the weight given to the point mass function at zero in \eqref{eq:prior}.

Another evident characteristic is that, regardless of the value of $\alpha$, as $d$ increases in magnitude, the shrinkage rule asymptotically converges to the value of $\tau$ when $d > 0$ and to $-\tau$ when $d<0$, indicating that wavelet coefficients with high magnitude are preserved, with the limiting value at the extremes of the raised cosine distribution support established by $\tau$. Hence, an empirical coefficient $d$ greater than $\tau$ in absolute value occurs due the presence of the noise $\epsilon$ and the shrinkage rule acts by taking the estimate $\hat{\theta}$ back to the support $(-\tau,\tau)$ of $\theta$.

\begin{figure}
    \centering
    \includegraphics[width=.8\linewidth]{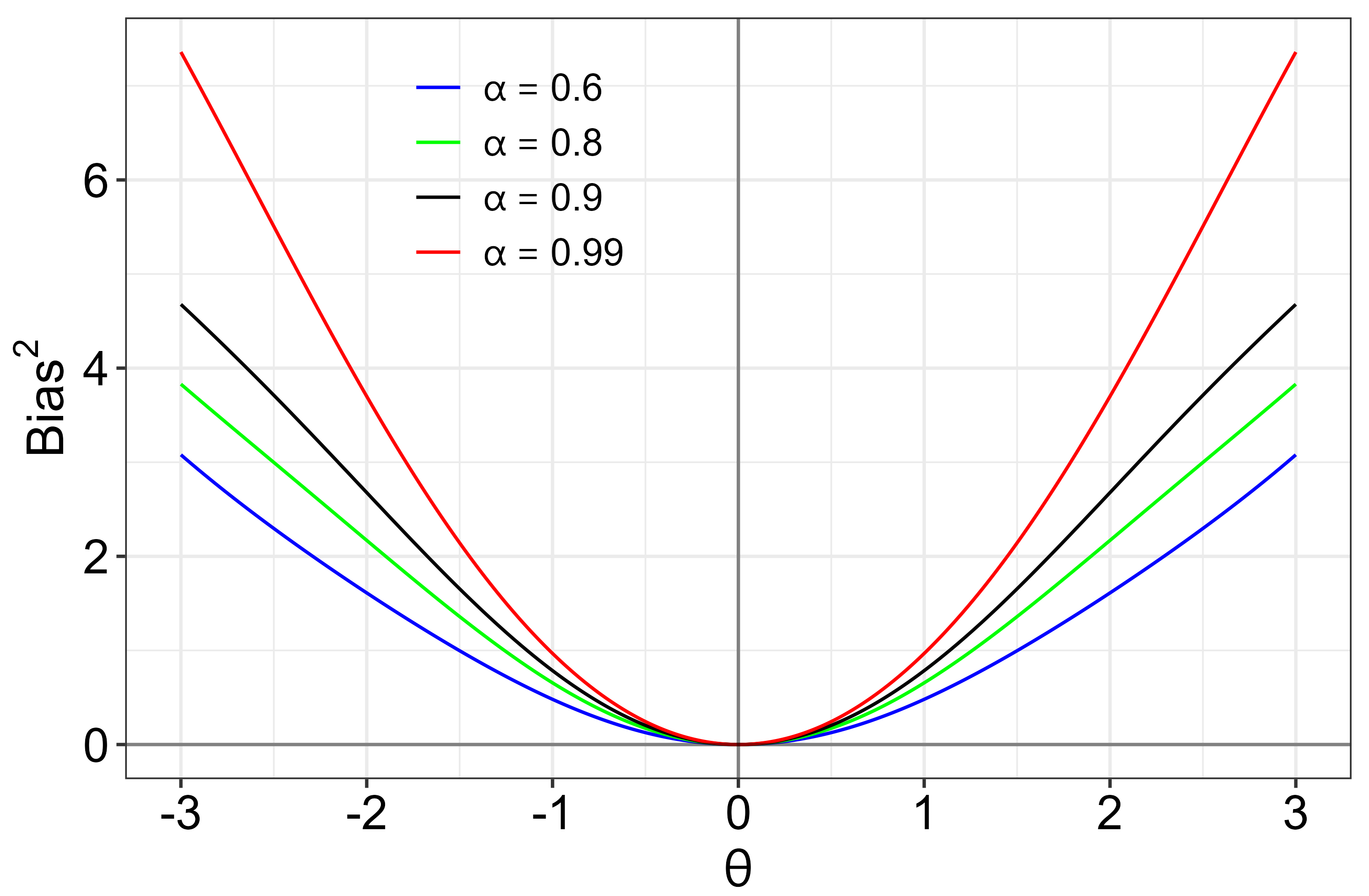} 
    \caption{Squared bias of the shrinkage rules under raised cosine prior for $\sigma = 1$, $\tau = 3$ and $\alpha \in \{0.6; 0.8; 0.9; 0.99\}$.}
    \label{bias:app}
\end{figure}

\begin{figure}
    \centering
    \includegraphics[width=.8\linewidth]{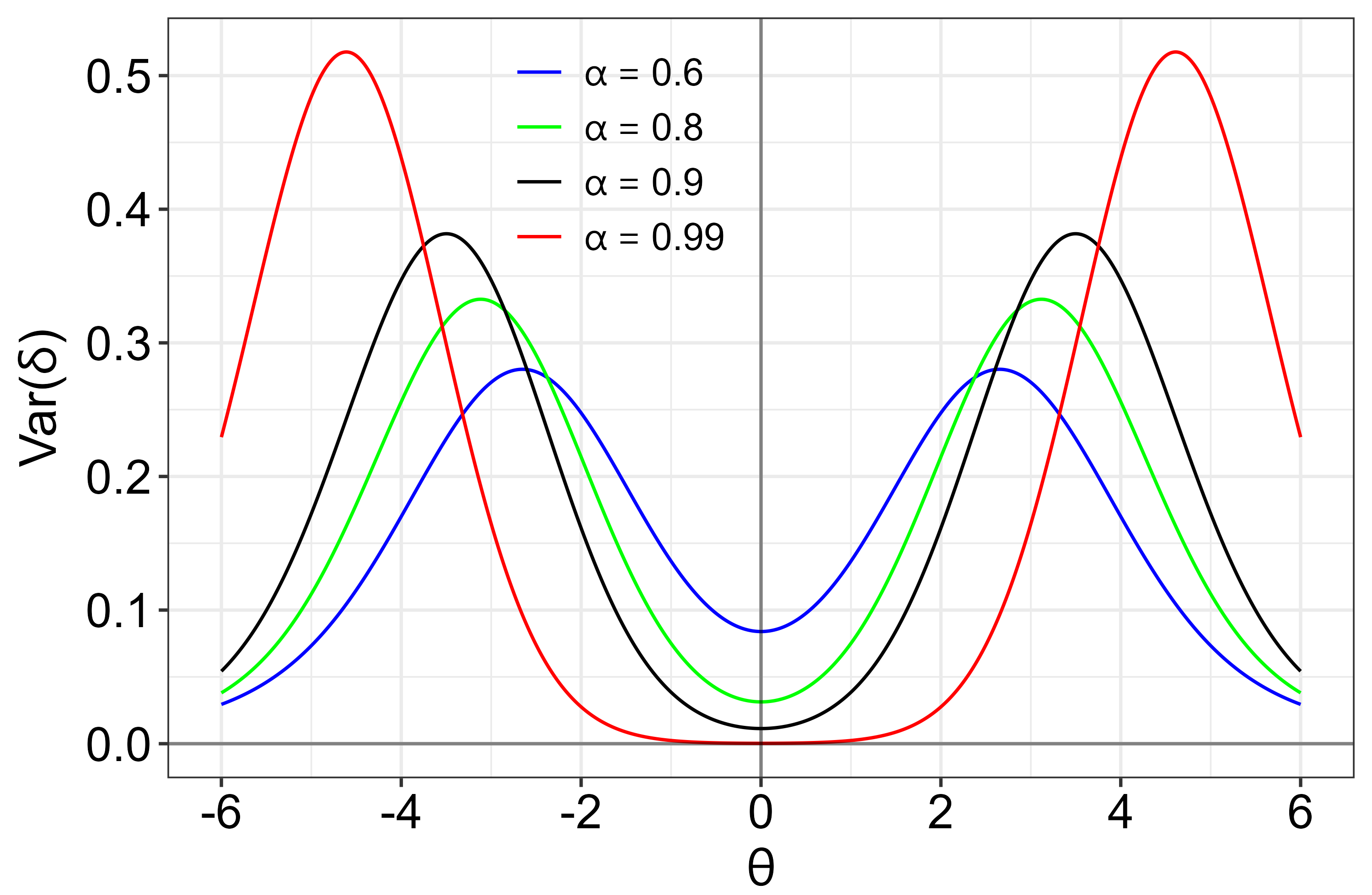} 
    \caption{Variance of the shrinkage rules under raised cosine prior for $\sigma = 1$, $\tau = 3$ and $\alpha \in \{0.6; 0.8; 0.9; 0.99\}$.}
    \label{var:app}
\end{figure}

\begin{figure}
    \centering
    \includegraphics[width=.8\linewidth]{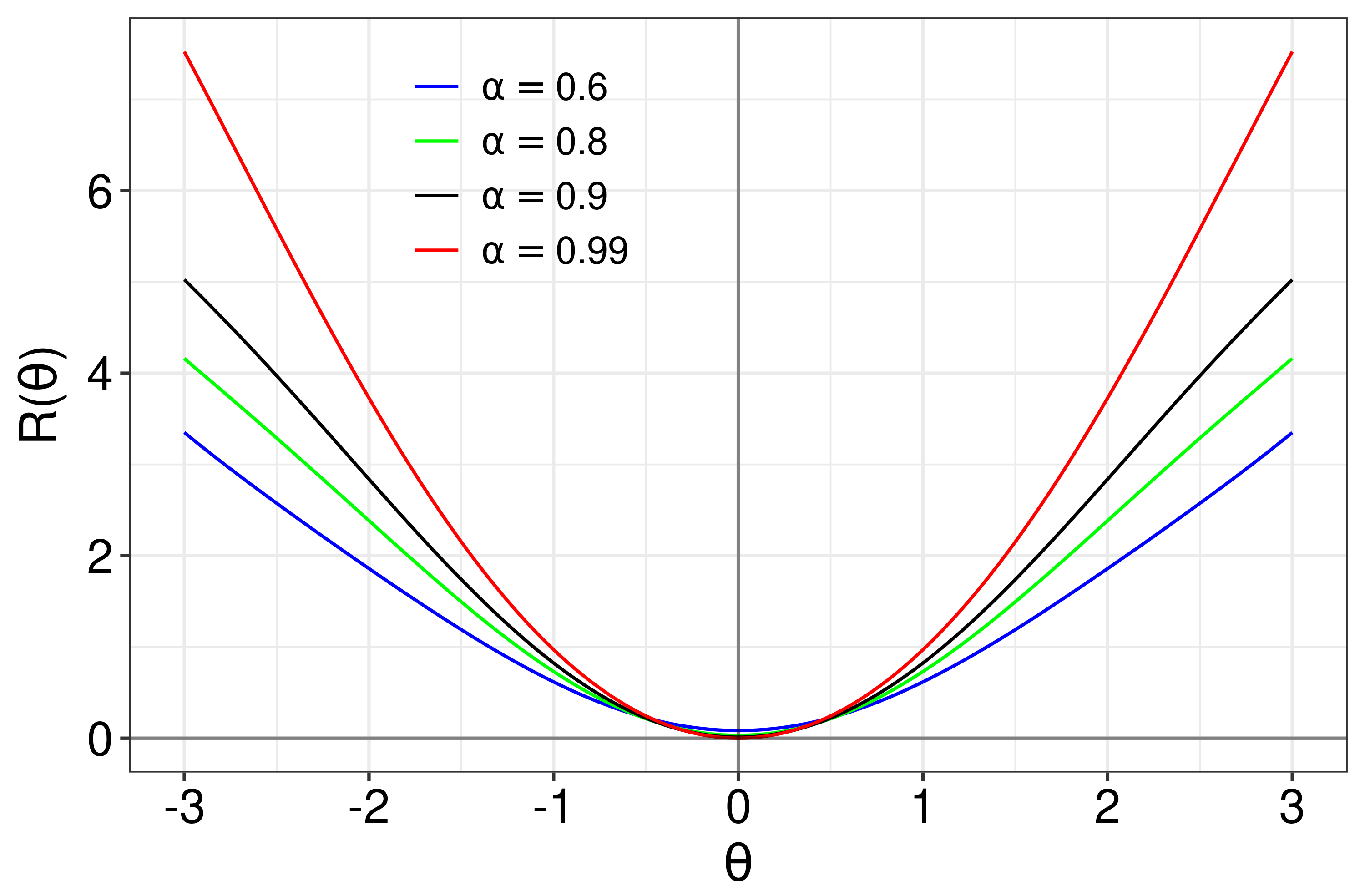} 
    \caption{Classical risk of the shrinkage rules under raised cosine prior for $\sigma = 1$, $\tau = 3$ and $\alpha \in \{0.6; 0.8; 0.9; 0.99\}$.}
    \label{clasrisk:app}
\end{figure}

Regarding the proposed rule for the specific cases in Figure \ref{rules1:app}, we provide plots to evaluate specific statistical properties of them. Figure \ref{bias:app} shows the squared bias of the rules. The bias is slight when $\theta$ is close to zero and increases for higher values of $\theta$ in absolute value. Moreover, rules with higher values of $\alpha$ have higher bias. Figure \ref{var:app} presents the variance of the rules, which we see is also small when $\theta$ is close to zero; it increases symmetrically until it reaches a peak and then decreases. Higher values of $\alpha$ are associated with a smaller variance for $\theta$ near zero, but a larger variance for $\theta$ far from zero. Thus, the rules have small bias and variance when $\theta$ is close to zero, which is interesting from the wavelet point of view since most of the wavelet coefficients are usually equal to zero. Figure \ref{clasrisk:app} presents the classical risk $R(\theta) = \mathbb{E}[L(\delta(d),\theta)]$ of the rules, which behaves the same as the squared bias, i.e, the risk increases as $\theta$ increases.  

Finally, the Bayesian risks $r = \mathbb{E}_{\pi}[ R^{\mathstrut} (\theta)] $ are provided in Table \ref{bayesrisk} for the shrinkage rules with the same range of values of $\alpha$ as the other propoerties discussed above, but also for $\tau \in \{1,2,3\}$. The results indicate that, due to the scale parameter characteristic of controlling the shape of the raised cosine distribution, higher values of $\tau$ result in greater Bayesian risk, as the distribution becomes less concentrated around zero. Similarly, for higher values of $\alpha$, the Bayesian risk is lower, which is consistent with the prior information regarding the sparsity of the wavelet coefficient vector.

\begin{table}[H]
\centering
\caption{Bayes risks of the shrinkage rules under raised cosine prior.}
\label{bayesrisk}
\begin{tabular}{|c||*{4}{c|}}\hline
\diagbox[width=4em, height=2.5em]{\large $\tau$}{\large $\alpha$} & $0.6$ & $0.8$ & $0.9$ & $0.99$ \\\hline\hline
\multicolumn{1}{|c||}{1} & 0.049 & 0.025 & 0.012 & 0.001 \\\hline
\multicolumn{1}{|c||}{2} & 0.171 & 0.093 & 0.049 & 0.005 \\\hline
\multicolumn{1}{|c||}{3} & 0.309 & 0.180 & 0.099 & 0.011 \\\hline
\end{tabular}
\end{table}

The performance of the Bayesian shrinkage rule depends on the choice of methods used to estimate the corresponding parameters and hyperparameters. Thus, a choice based on known prior characteristics is essential to determine the degree of shrinkage of the wavelet coefficients, ensuring the structural integrity of the signal of interest. The rule proposed in this work under the raised cosine prior requires the estimation of the parameter $\sigma$ and the elicitation of the hyperparameters $\alpha$ and $\tau$. 

For the estimation of $\sigma$, we use the standard robust estimate $\hat{\sigma}$ proposed by \cite{DJ-1994},
\begin{equation} \label{eq:est-sigma}
    \hat{\sigma} = \frac{\text{median} \{ |d_{J-1, k}|, k = 0, \dots, 2^{J-1} \}}{0.6745},
\end{equation}

\noindent where the constant $0.6745$ represents the $75^{\text{th}}$ percentile of the standard normal distribution.

The elicitation of $\alpha$ can be done according to \citet{angelini-vidakovic-2004}, 
\begin{equation}\label{eq:alpha}
\alpha = \alpha(j) = 1 - \frac{1}{(j-J_{0}+1)^\gamma},
\end{equation}
where $J_ 0 \leq j \leq J-1$, $J_0$ is the primary resolution level, $J$ is the number of resolution levels, $J=\log_{2}(n)$ and $\gamma > 0$. They also suggested that in the absence of additional information, $\gamma = 2$ can be adopted. We suggest the elicitation of $\tau$ to be done according to an adaptation of the level-dependent proposal of \cite{alex-beta} to the hyperparameter of the support of the beta prior distribution to $\theta$ on $(-\tau,\tau)$,
\begin{equation} \label{eq:tau}
    \tau = \underset{j,k}{\max} \{ |d_{j,k}|\}.
\end{equation}

\section{Simulation studies}
\begin{figure}
    \centering
    \includegraphics[width=.8\linewidth]{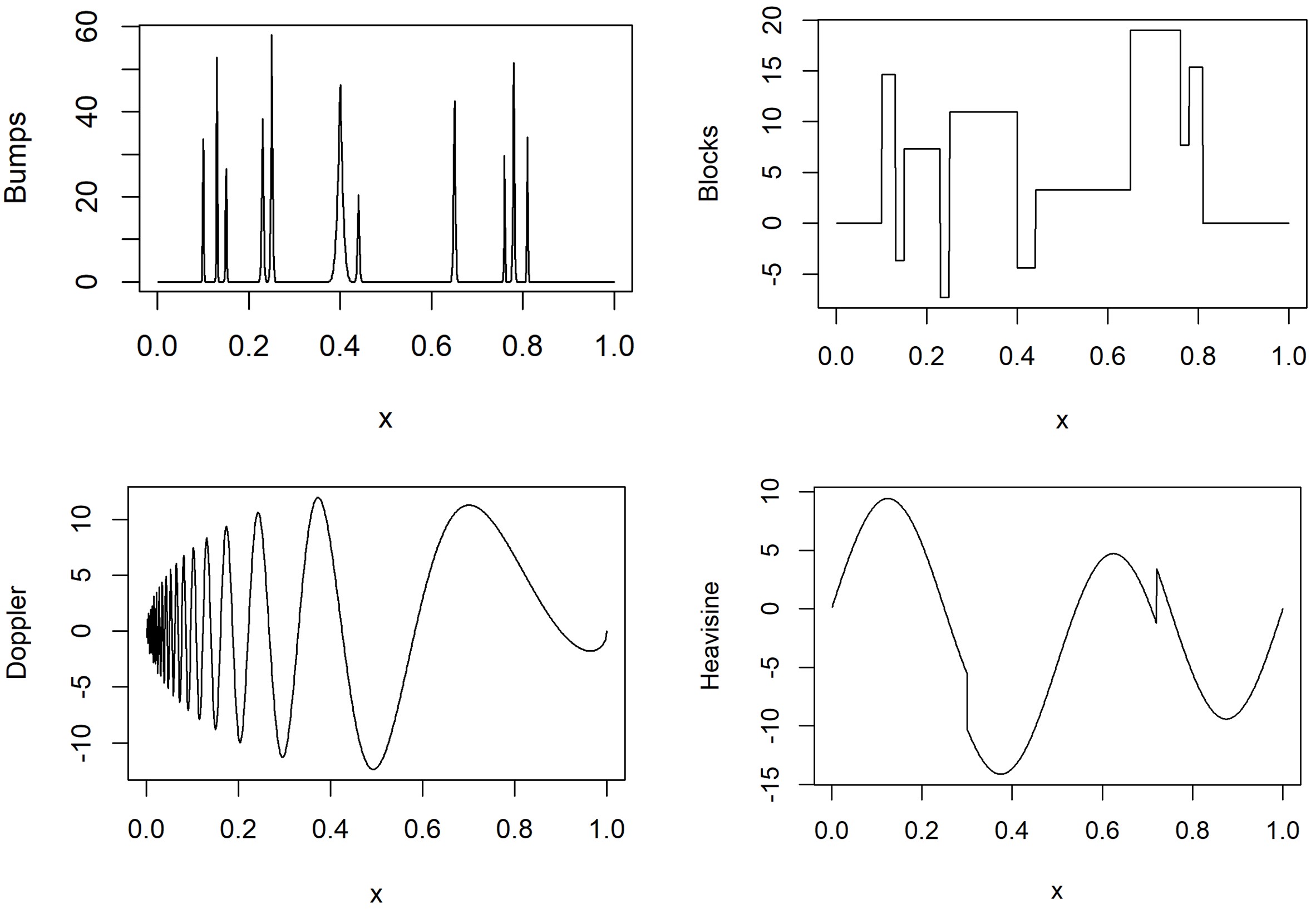} 
    \caption{Donoho and Johnstone test functions.}
    \label{dj_functions:app}
\end{figure}

\indent In order to evaluate and compare the performance of the proposed estimator with other Bayesian shrinkage estimators as well as with classical threshold estimators, simulations were conducted in different predefined scenarios. These scenarios include variations in sample size $n = 128, 512, 1024$ and $2048$, signal-to-noise ratio (SNR), $\text{SNR} = 1, 3, 6$ and $9$, and underlying function $f$ in \eqref{time_model}, in which the so called Donoho-Johnstone (D-J) test functions were considered. These functions, called Bumps, Blocks, Doppler, and Heavisine, are widely used in the wavelet literature to assess the effectiveness of estimation methods in signal recovery problems. Their ability to represent different regularity patterns and discontinuities makes them particularly relevant for real-world signal applications, see \cite{DJ-1995} for more details about the D-J test functions. Further, the formulas defining each of these functions can be found in Table \ref{func}, while their corresponding graphs are shown in Figure \ref{dj_functions:app}.

\begin{table}[H]
\caption{Definitions of the Donoho and Johnstone functions.}
\label{func}
\centering
\renewcommand{\arraystretch}{1.8}
\begin{tabular}{>{\centering\arraybackslash}p{2.5cm} >{\centering\arraybackslash}p{13cm}}
\hline
\textbf{Function} & \textbf{Definition} \\
\hline
\multirow{5}{*}{Bumps} & $f(x) = \sum h_j K\left(\frac{x - x_j}{w_j}\right), \hspace{0.3cm} x \in [0,1]$ \\
 & $K(x) = (1 + |x|)^{-4}$ \\
 & $\{x_j\} = (0.1, 0.13, 0.15, 0.23, 0.25, 0.40, 0.44, 0.65, 0.76, 0.78, 0.81)$ \\
 & $\{h_j\} = (4, 5, 3, 4, 5, 4.2, 2.1, 4.3, 3.1, 5.1, 4.2)$ \\
 & $\{w_j\} = (0.005, 0.005, 0.006, 0.01, 0.01, 0.03, 0.01, 0.01, 0.005, 0.008, 0.005)$ \\
\hline
\multirow{4}{*}{Blocks} & $f(x) = \sum h_j K(x - x_j), \hspace{0.3cm} x \in [0,1]$ \\
 & $K(x) = (1 + \text{sgn}(x))/2$ \\
 & $\{x_j\} = x_\text{Bumps}$ \\
 & $\{h_j\} = (4, -5, 3, -4, 5, -4.2, 2.1, 4.3, -3.1, 2.1, -4.2)$ \\
\hline
Doppler & $f(x) = \sqrt{x(1 - x)}\sin \left( \frac{2\pi (1 + 0.05)}{x + 0.05} \right), \hspace{0.3cm} x \in [0,1]$ \\
\hline
Heavisine & $f(x) = 4 \sin(4 \pi x) - \text{sgn}(x - 0.3) - \text{sgn}(0.72 - x), \hspace{0.3cm} x \in [0,1]$ \\
\hline
\end{tabular}
\end{table}

We compared the performance of the proposed rule with the soft thresholding rule $\eta_S(d)$ of \cite{DJ-1994} given by
\begin{equation}\label{softrule} 
\eta_S(d) = \begin{cases} 
 0, & \text{if $|d| \leq \lambda$} \\  
 \mathrm{sgn}(d)(|d| - \lambda), & \text{if $|d|>\lambda$,}   
 \end{cases} 
\end{equation}
where $\lambda > 0$ and $\mathrm{sgn}(\cdot)$ is the sign function,
with policy to choose the threshold value $\lambda$ in \eqref{softrule} given by the classical methods: Universal threshold proposed by \cite{DJ-1994}, False Discovery Rate (FDR) by \cite{abra-1996}, Cross Validation (CV) by \cite{nason-1996} and Stein Unbiased Risk Estimator (SURE) by \cite{DJ-1995}. We also included the Bayesian shrinkage rule based on the beta prior of \cite{alex-beta}, which the prior distribution to $\theta$ has the form \eqref{eq:prior} but with $g(\theta)$ being a symmetrically around zero beta probability density function on a bounded support $(-m, m)$, i.e,
\begin{equation} \label{betaprior}
    g(\theta;m,a) = \frac{(m^2 - \theta^2)^{(a-1)}}{(2m)^{(2a-1)}B(a,a)}\mathbb{I}_{(-m, ~ m)} (\theta),
\end{equation}
where $m>0$, $a > 0$ and $B(\cdot,\cdot)$ is the beta function. We considered the beta shrinkage rules with $a = 1$ and $a = 5$ in \eqref{betaprior}, as these hyperparameter values yielded the best overall performance in the simulation studies conducted by the authors, as shown in \cite{alex-beta}. Further, when $a=1$, the beta shrinkage rule is equivalent to the shrinkage rule under a uniform prior of \cite{angelini-vidakovic-2004}. Furthermore, the hyperparameter $m$, associated with the beta distribution, and the hyperparameter $\tau$, related to the raised cosine distribution, were obtained according to \eqref{eq:tau}. Although $\alpha$ could be elicited using \eqref{eq:alpha}, we fixed $\alpha = 0.9$ for all Bayesian rules and resolution levels.

For each test function, sample size, and SNR, samples were generated according to the model in Equation \eqref{time_model}. The standard deviation $\sigma$ of the errors was obtained according to the SNR, i.e, $\sigma = \mathrm{SD}(f)/\mathrm{SNR}$, where $\mathrm{SD}(f)$ is the standard deviation of the signal $f$ (for the D-J test functions, $\mathrm{SD}(f) = 7$). After the data generation, the DWT was applied to the noisy data vector, using the Daubechies wavelet basis with 10 vanishing moments (Daub10). This application yields a vector of empirical coefficients, which represent the contributions from different scales and locations of the original signal, thereby situating them in the wavelet domain. The next step consists of applying the shrinkage and thresholding rules to the empirical wavelet coefficients to estimate the wavelet coefficients. After this step, the inverse discrete wavelet transform (IDWT) is performed, allowing the recovery of the data in its original domain and the reconstruction of the estimated signal. We used the R package \texttt{WaveThresh} from \cite{wavethresh} to perform the stages of data generation, DWT/IDWT, and soft thresholding application.

To measure the performance of the rules, we calculated for each replication $r$ and for each rule, the mean squared error (MSE) and the median absolute error (MAE) given by
\begin{equation} \nonumber
 \mathrm{MSE}_r = \frac{1}{n} \sum_{i=1}^{n} \left( \hat{f}^{(r)}(x_i) - f(x_i) \right)^2, 
 \end{equation}

\begin{equation} \nonumber
    \mathrm{MAE}_r = \text{median} \left( \left| \hat{f}^{(r)}(x_i) - f(x_i) \right| \right),
\end{equation}

\noindent where \(\hat{f}^{(r)}(x_i)\) denotes the estimate of $f(x_i)$, for \(i = 1, \dots, n\). Then, we finally obtained the averaged mean squared error (AMSE) and averaged median absolute error (AMAE) for each rule, i.e,
\begin{equation} \nonumber
\mathrm{AMSE} = \frac{1}{R} \sum_{r=1}^{R} \mathrm{MSE}_r,  \end{equation}

\begin{equation} \nonumber 
\mathrm{AMAE} = \frac{1}{R} \sum_{r=1}^{R} \mathrm{MAE}_r,  
\end{equation}

\noindent where $R$ represents the number of replications performed. In our case, $R = 200$. Strictly restricted to these measures, a rule performed well if it had smaller values of AMSE and AMAE than its competitors. For the remainder of this section, we discuss the results of the AMSE. The tables with the AMAE are available in the Supplementary Material.

Tables \ref{amse1}, \ref{amse2}, \ref{amse3}, and \ref{amse4} show the AMSE with the standard deviation of the MSEs for $\mathrm{SNR} = 1, 3, 6$ and $9$, respectively. As expected, the AMSE of the methods decreases as the sample size or the SNR increases, because a larger sample provides a more accurate representation of the underlying signal, and a larger SNR means a weaker presence of noise in the dataset. It is also noted that, in general, Bayesian rules perform better or yield lower AMSE values in scenarios with lower signal-to-noise ratios, such as when SNR = 1. In this case, the Bayesian rules outperformed all classical methods, with the shrinkage rule under raised cosine prior achieving the best performance for the test functions Bumps and Doppler when $n = 512$ and $2048$. 

The proposed shrinkage rule under the raised cosine prior worked well in general, particularly for low SNR. For $\mathrm{SNR} = 1$, it had the best performance for Bumps as the underlying function when $n = 512$ and $2048$, and also for the Doppler function when $n = 512$. Although the beta shrinkage rule with $a=1$ outperformed the other rules for Blocks and Heavisine functions, the proposed rule was the second-best, with AMSEs very close to those of the beta rule. For instance, its AMSE was $8.564$ for the Blocks function and $n = 1024$, compared to $8.406$ for the beta rule in the same scenario. On the other hand, for $\mathrm{SNR} = 3$, the proposed rule was the best in all sample size scenarios for the Doppler function,  and for Bumps and Heavisine functions when $n = 128$. 

For $\mathrm{SNR} = 6$ and $9$, the soft thresholding rule with CV and SURE policies had the best overall performance. However, the proposed rule was also competitive. For instance, its AMSE was equal to $0.804$ for the Bumps function when $\mathrm{SNR} = 9$ and $n = 512$ against $0.722$ of the thresholding rule with SURE policy. A similar overall behavior was observed for the AMAE measure.

\begin{figure}
    \centering
    \includegraphics[width=.8\linewidth]{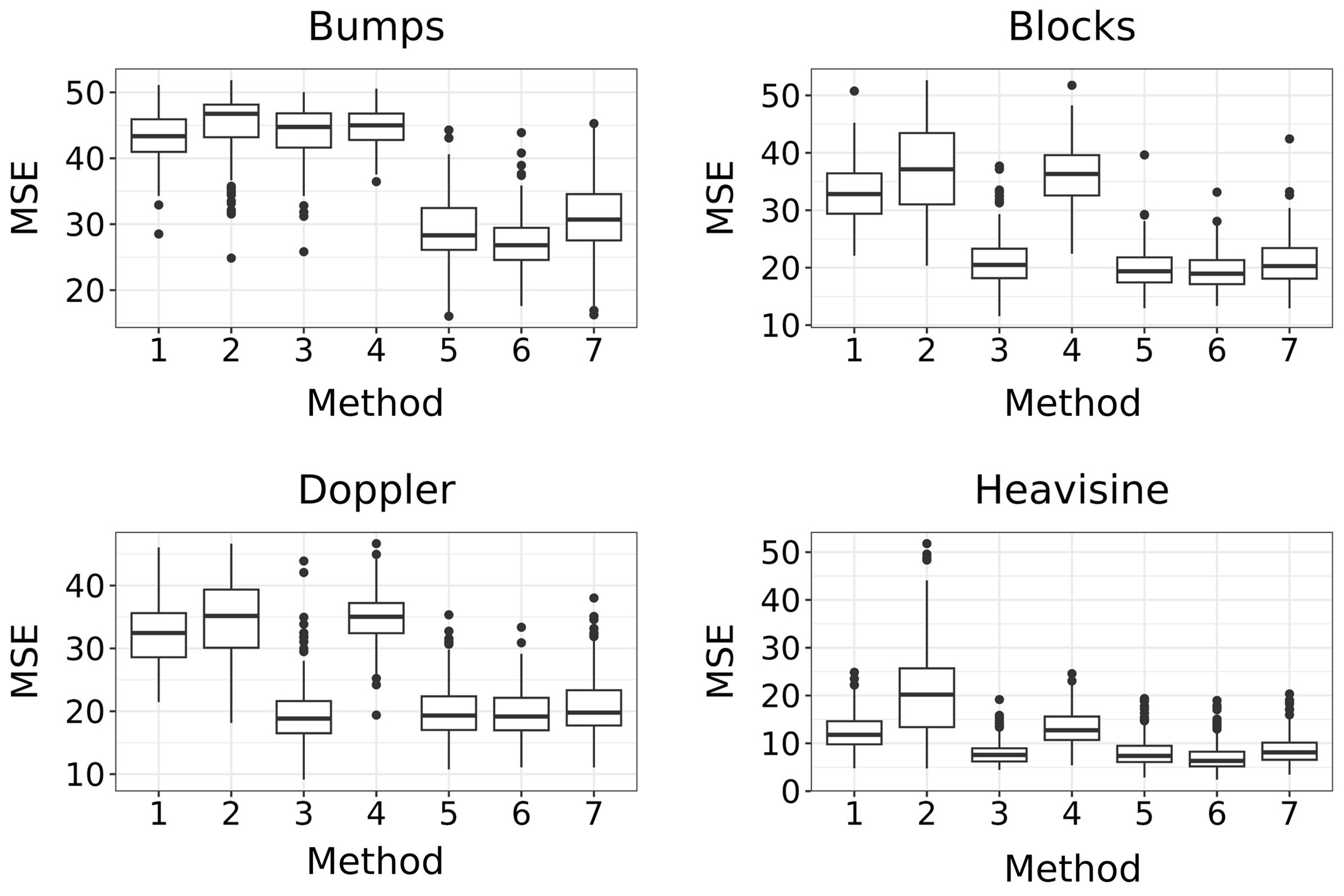} 
    \caption{Boxplots of the MSE for $n = 128$ and $\mathrm{SNR} = 1$. The corresponding rules are: 1-Universal, 2-FDR, 3-CV, 4-SURE, 5-Raised cosine, 6-Beta(1,1) and 7-Beta(5,5).}
    \label{bp:app}
\end{figure}

Figure \ref{bp:app} shows the boxplots of the MSE for $n = 128$ and $\mathrm{SNR} = 1$. The behavior of the Bayesian rules was similar in terms of dispersion. It outperformed the soft thresholding rule with universal, FDR, CV, and SURE policies for Bumps, Blocks, and Doppler functions. The boxplots of the MSE for $\mathrm{SNR} = 1$ and $n = 512, 1024$ and $2048$ as the boxplots of the AMAE for all scenarios of $\mathrm{SNR} = 1$, are available in the Supplementary Material.

\begin{table}[H]
\caption{AMSE (standard deviation) of shrinkage and thresholding rules for the D-J test functions with SNR = 1.}
\label{amse1}
\centering

\vspace{0.3cm}

\begin{tabular}{cccccc}
\hline
Método    &  n    & Bumps          & Blocks         & Doppler        & Heavisine \\ \hline
          \multirow{4}{*}{\centering Universal} 
          & 128   & 43.098 (3.680) & 32.905 (5.158) & 32.374 (4.800) & 12.364 (3.797)  \\ 
          & 512   & 41.372 (2.250) & 20.563 (1.972) & 20.671 (2.277) & 7.218 (1.458)   \\ 
          & 1024  & 33.620 (2.060) & 16.110 (1.353) & 14.927 (1.451) & 4.807 (0.884)   \\ 
          & 2048  & 25.794 (1.414) & 12.559 (0.771) & 10.266 (0.825) & 3.059 (0.446)   \\  \hline
          \multirow{4}{*}{\centering FDR}
          & 128   & 44.976 (4.561) & 37.112 (7.540) & 34.735 (6.383) & 21.028 (9.426)  \\ 
          & 512   & 39.028 (3.283) & 20.406 (2.727) & 19.818 (2.895) & 9.084 (3.713)   \\
          & 1024  & 29.393 (2.424) & 15.388 (1.675) & 13.563 (1.559) & 6.276 (1.707)  \\
          & 2048  & 21.245 (1.509) & 11.626 (0.907) & 9.150  (1.037) & 4.404 (1.717)   \\ \hline
          \multirow{4}{*}{\centering CV}
          & 128   & 43.867 (3.794) & 21.098 (4.315) & \textbf{19.560} (4.952) & 7.891 (2.378)   \\ 
          & 512   & 24.207 (3.168) & 11.607 (1.254) & 9.969  (1.258) & 4.060 (0.942)   \\ 
          & 1024  & 15.607 (1.211) & 9.136  (0.771) & 6.850  (0.707) & 2.680 (0.554)   \\ 
          & 2048  & 10.872 (0.761) & 6.958  (0.511) & 4.672  (0.463) & 1.770 (0.279)   \\ \hline
          \multirow{4}{*}{\centering SURE}
          & 128   & 44.747 (2.746) & 36.283 (5.004) & 35.028 (4.212) & 13.335 (3.614)  \\ 
          & 512   & 43.712 (1.827) & 21.880 (1.852) & 22.021 (2.070) & 7.315 (1.529)   \\ 
          & 1024  & 36.421 (1.735) & 16.736 (1.840) & 15.445 (1.853) & 4.463 (1.277)  \\
          & 2048  & 26.306 (5.026) & 9.787  (3.264) & 8.471  (3.107) & 1.943 (0.522)   \\ \hline
          \multirow{4}{*}{\centering Raised Cosine}
          & 128   & 29.108 (5.218) & 19.871 (3.567) & 19.882 (4.386) & 8.122 (3.096)    \\ 
          & 512   & \textbf{21.224} (2.175) & 11.063 (1.454) & \textbf{8.976} (1.376) & 3.816 (1.168) \\ 
              & 1024  & 14.795 (1.161) & 8.564 (0.805)  & 5.737 (0.836) & 2.530 (0.694) \\ 
              & 2048  & \textbf{9.660}  (0.740) & 6.490 (0.565)  & 3.768 (0.566) & 1.822 (0.481) \\  \hline
              \multirow{4}{*}{\centering Beta$(1,1)$}
              & 128   & \textbf{27.248} (4.293) & \textbf{19.483} (3.192)  & 19.655 (3.780)  & \textbf{7.133} (2.985)  \\ 
              & 512   & 21.759 (1.982) & \textbf{11.042} (1.416) & 8.978 (1.419) & \textbf{3.297} (1.082) \\
              & 1024  & 15.551 (1.196) & \textbf{8.406}  (0.749) & \textbf{5.463} (0.768) & \textbf{2.029} (0.637)  \\ 
              & 2048  & 9.887  (0.723) & \textbf{6.379}  (0.513) & \textbf{3.428} (0.496) & \textbf{1.458} (0.412) \\ \hline  
             \multirow{4}{*}{\centering Beta$(5,5)$}
             & 128   & 30.952 (5.666)  & 20.823 (4.059)  & 20.823 (4.941)  &  8.687 (3.116)   \\ 
             & 512   & 21.994 (2.542)          & 11.238 (1.454) & 9.134 (1.377) & 4.124 (1.235)  \\ 
             & 1024  & \textbf{14.781} (1.185) & 8.721  (0.844) & 5.929 (0.870) & 2.744 (0.764)\\ 
              & 2048  & 9.667 (0.746)           & 6.610  (0.579) & 3.956 (0.563) & 2.032 (0.528) \\ \hline                
\end{tabular}
\end{table}

\begin{table}[H]
\caption{AMSE (standard deviation) of shrinkage and thresholding rules for the D-J test functions with SNR = 3.}
\label{amse2}
\centering

\vspace{0.3cm}

\begin{tabular}{cccccc}
\hline
Método    &  n    & Bumps          & Blocks        & Doppler       & Heavisine    \\ \hline
          \multirow{4}{*}{\centering Universal} 
          & 128   & 24.538 (3.885) & 16.257 (2.390) & 12.764 (2.459) & 3.121 (0.738) \\ 
          & 512   & 17.247 (1.722) & 8.058 (0.631) & 5.013 (0.628) & 1.506 (0.225) \\ 
          & 1024  & 10.704 (0.868) & 6.055 (0.427) & 3.204 (0.306) & 1.064 (0.123)  \\ 
          & 2048  & 6.631  (0.387) & 4.348 (0.241) & 2.082 (0.160) & 0.753 (0.070)  \\  \hline
          \multirow{4}{*}{\centering FDR} 
          & 128   & 21.905 (4.592) & 14.408 (2.765) & 11.410 (2.876) & 3.871 (1.989) \\ 
          & 512   & 12.028 (1.450) & 6.673 (0.698) & 4.061 (0.644) & 1.996 (1.066)   \\
          & 1024  & 7.068  (0.686) & 4.741 (0.421) & 2.585 (0.321) & 1.188 (0.780)  \\
          & 2048  & 4.406  (0.300) & 3.282 (0.242) & 1.649 (0.161) & 0.752 (0.108)  \\ \hline
          \multirow{4}{*}{\centering CV} 
          & 128   & 32.832 (3.247) & 7.116 (1.329) & 4.630 (0.785) & 1.720 (0.393)  \\ 
          & 512   & 9.424  (0.965) & \textbf{2.902} (0.242) & 1.655 (0.190) & \textbf{0.825} (0.119)  \\ 
          & 1024  & 3.285  (0.227) & \textbf{2.154} (0.170) & 1.158 (0.105) & \textbf{0.586} (0.073)  \\ 
          & 2048  & 1.867  (0.108) & \textbf{1.497} (0.099) & 0.765 (0.068) & \textbf{0.415} (0.041)  \\ \hline
          \multirow{4}{*}{\centering SURE} 
          & 128   & 18.679 (13.162) & 7.168 (3.889) & 4.574 (1.138) & 1.769 (0.420) \\ 
          & 512   & 4.649 (0.690) & 3.303 (0.489) & 1.767 (0.242) & 0.871 (0.128)  \\ 
          & 1024  & 2.982 (0.338) & 2.325 (0.243) & 1.213 (0.129) & 0.613 (0.087)  \\
          & 2048  & 1.990 (0.159) & 1.593 (0.129) & 0.784 (0.074) & 0.428 (0.049)  \\ \hline
          \multirow{4}{*}{\centering Raised Cosine} 
          & 128   & \textbf{8.779} (1.752) & 6.780 (1.299) & \textbf{4.440} (0.907) & \textbf{1.588} (0.449)   \\ 
          & 512   & 4.250 (0.448) & 3.613 (0.297) & \textbf{1.347} (0.207) & 0.899 (0.302)  \\ 
          & 1024  & 2.506 (0.205) & 2.471 (0.220) & \textbf{0.951} (0.132) & 0.842 (0.448)\\ 
              & 2048  & 1.599 (0.108) & 1.713 (0.181) & \textbf{0.708} (0.125) & 0.955 (0.728) \\  \hline
              \multirow{4}{*}{\centering Beta$(1,1)$}
              & 128   & 9.300 (1.685) & 7.417 (1.416) & 4.770 (0.956) & 1.593 (0.418)   \\ 
              & 512   & 4.734 (0.516) & 3.914 (0.329) & 1.404 (0.219) & 0.996 (0.498) \\
              & 1024  & 2.726 (0.230) & 2.692 (0.261) & 1.008 (0.164) & 0.942 (0.613)  \\ 
              & 2048  & 1.693 (0.106) & 1.856 (0.258) & 0.761 (0.173) & 0.944 (0.714) \\ \hline  
             \multirow{4}{*}{\centering Beta$(5,5)$}
             & 128   &  9.058 (1.969) & \textbf{6.770} (1.331) & 4.452 (0.929) & 1.615 (0.427)    \\ 
             & 512   & \textbf{4.183} (0.442) & 3.537 (0.310) & 1.352 (0.218) & 0.948 (0.369) \\ 
             & 1024  & \textbf{2.453} (0.203) & 2.433 (0.210) & 0.977 (0.149) & 0.831 (0.429) \\ 
              & 2048  & \textbf{1.571} (0.100) & 1.704 (0.232) & 0.723 (0.171) & 0.775 (0.426) \\ \hline                
\end{tabular}
\end{table}

\begin{table}[H]
\caption{AMSE (standard deviation) of shrinkage and thresholding rules for the D-J test functions with SNR = 6.}
\label{amse3}
\centering

\vspace{0.3cm}

\begin{tabular}{cccccc}
\hline
Método    &  n    & Bumps         & Blocks        & Doppler       & Heavisine     \\ \hline
          \multirow{4}{*}{\centering Universal} 
          & 128   & 14.982 (2.387) & 8.797 (1.725) & 5.345 (1.176) & 1.311 (0.281)  \\ 
          & 512   & 7.151 (0.737) & 4.108 (0.388) & 1.742 (0.220) & 0.678 (0.081) \\ 
          & 1024  & 3.899 (0.345) & 2.713 (0.215) & 1.136 (0.104) & 0.477 (0.041)  \\ 
          & 2048  & 2.370 (0.137) & 1.794 (0.109) & 0.691 (0.051) & 0.334 (0.026)  \\  \hline
          \multirow{4}{*}{\centering FDR} 
          & 128   & 11.998 (2.508) & 6.640 (1.550) & 4.097 (1.040) & 1.465 (0.609) \\ 
          & 512   & 4.361 (0.524) & 2.799 (0.328) & 1.330 (0.218) & 0.638 (0.095)  \\
          & 1024  & 2.316 (0.224) & 1.787 (0.171) & 0.853 (0.096) & 0.434 (0.051)  \\
          & 2048  & 1.447 (0.098) & 1.181 (0.087) & 0.511 (0.051) & 0.294 (0.031) \\ \hline
          \multirow{4}{*}{\centering CV} 
          & 128   & 25.998 (2.424) & 4.797 (0.619) & 2.779 (0.342) & 0.757 (0.174) \\ 
          & 512   & 7.736 (0.695) & 1.214 (0.104) & \textbf{0.539} (0.063) & \textbf{0.314} (0.040)  \\ 
          & 1024  & 1.852 (0.138) & 0.806 (0.061) & \textbf{0.362} (0.031) & \textbf{0.221} (0.022)  \\ 
          & 2048  & 0.666 (0.037) & \textbf{0.491} (0.031) & \textbf{0.223} (0.018) & \textbf{0.153} (0.013)  \\ \hline
          \multirow{4}{*}{\centering SURE} 
          & 128   & 4.908 (1.679) & \textbf{2.371} (0.560) & \textbf{1.397} (0.363) & \textbf{0.680} (0.165)  \\ 
          & 512   & 1.470 (0.206) & \textbf{1.114} (0.174) & 0.545 (0.075) & 0.335 (0.049) \\ 
          & 1024  & 0.890 (0.100) & \textbf{0.751} (0.078) & 0.373 (0.036) & 0.230 (0.025)  \\
          & 2048  & \textbf{0.604} (0.050) & 0.513 (0.041)          & 0.230 (0.019) & 0.157 (0.016)  \\ \hline
              \multirow{4}{*}{\centering Raised Cosine} 
              & 128   & 4.392 (0.822)  & 2.533 (0.637)  & 1.457 (0.372)  & 0.867 (0.326) \\ 
              & 512   & 1.458 (0.147)  & 1.471 (0.197)  & 0.601 (0.137)  & 0.822 (0.652) \\ 
              & 1024  & 0.901 (0.081)  & 1.112 (0.224)  & 0.578 (0.164)  & 1.048 (0.973) \\ 
              & 2048  & 0.752 (0.093)  & 0.990 (0.291)  & 0.562 (0.233)  & 1.431 (1.596) \\  \hline
              \multirow{4}{*}{\centering Beta$(1,1)$} 
              & 128   & 4.891 (0.875)  & 2.892 (0.735)  & 1.620 (0.414)  & 0.991 (0.388)  \\ 
              & 512   & 1.628 (0.166)  & 1.654 (0.249)  & 0.657 (0.153)  & 1.072 (1.114) \\ 
              & 1024  & 0.974 (0.097)  & 1.255 (0.329)  & 0.649 (0.243)  & 1.138 (1.143) \\ 
              & 2048  & 0.782 (0.089)  & 1.096 (0.430)  & 0.648 (0.323)  & 1.611 (1.595) \\ \hline  
              \multirow{4}{*}{\centering Beta$(5,5)$}
              & 128   & \textbf{4.366} (0.863)  & 2.481 (0.636)  & 1.431 (0.377)  & 0.888 (0.338)  \\ 
              & 512   & \textbf{1.419} (0.141)  & 1.441 (0.204)  & 0.601 (0.130)  & 0.823 (0.599) \\ 
              & 1024  & \textbf{0.882} (0.086)  & 1.077 (0.213)  & 0.572 (0.173)  & 0.886 (0.698) \\ 
              & 2048  & 0.726 (0.071)  & 1.032 (0.399)  & 0.602 (0.282)  & 1.085 (1.404) \\ \hline                
\end{tabular}
\end{table}

\vspace{0.5cm}

\begin{table}[H]
\caption{AMSE (standard deviation) of shrinkage and thresholding rules for the D-J test functions with SNR = 9.}
\label{amse4}
\centering

\vspace{0.3cm}

\begin{tabular}{cccccc}
\hline
Método    &  n    & Bumps         & Blocks        & Doppler       & Heavisine \\ \hline
          \multirow{4}{*}{\centering Universal}
          & 128   & 11.403 (1.797) & 5.585 (1.131) & 2.881 (0.700) & 0.839 (0.166) \\ 
          & 512   & 3.999 (0.448) & 2.418 (0.261) & 0.927 (0.117) & 0.424 (0.048) \\ 
          & 1024  & 2.066 (0.189) & 1.515 (0.127) & 0.604 (0.056) & 0.290 (0.025)  \\ 
          & 2048  & 1.259 (0.074) & 0.988 (0.061) & 0.346 (0.025) & 0.197 (0.014)  \\  \hline
          \multirow{4}{*}{\centering FDR}
          & 128   & 8.607 (1.706) & 3.962 (0.938) & 2.165 (0.590) & 0.838 (0.252)  \\ 
          & 512   & 2.275 (0.290) & 1.501 (0.194) & 0.680 (0.101) & 0.372 (0.054)  \\
          & 1024  & 1.161 (0.117) & 0.926 (0.090) & 0.437 (0.051) & 0.247 (0.027)  \\
          & 2048  & 0.729 (0.050) & 0.612 (0.045) & 0.250 (0.023) & 0.165 (0.018)  \\ \hline
          \multirow{4}{*}{\centering CV}
          & 128   & 24.527 (2.032) & 4.369 (0.456) & 2.446 (0.229) & 0.518 (0.122)   \\ 
          & 512   & 6.981 (0.402) & 0.866 (0.068) & 0.310 (0.033) & \textbf{0.168} (0.021) \\ 
          & 1024  & 1.634 (0.115) & 0.534 (0.039) & 0.189 (0.016) & \textbf{0.117} (0.012)  \\ 
          & 2048  & 0.442 (0.023) & 0.265 (0.017) & \textbf{0.107} (0.008) & \textbf{0.080} (0.006)  \\ \hline
          \multirow{4}{*}{\centering SURE}
          & 128   & 3.978 (1.243) & 1.405 (0.427) & \textbf{0.666} (0.167) & \textbf{0.397} (0.093)     \\ 
          & 512   & \textbf{0.722} (0.105) & \textbf{0.544} (0.084) & \textbf{0.265} (0.037) & 0.179 (0.025)  \\ 
          & 1024  & \textbf{0.428} (0.047) & \textbf{0.367} (0.039) & \textbf{0.182} (0.017) & 0.122 (0.013)  \\
          & 2048  & \textbf{0.290} (0.023) & \textbf{0.252} (0.019) & 0.111 (0.009)          & 0.082 (0.007)  \\ \hline
          \multirow{4}{*}{\centering Raised Cosine}
          & 128   &  3.064 (0.586)     & 1.305 (0.296)  & 0.698 (0.176)  & 0.797 (0.563)  \\ 
          & 512       & 0.804 (0.081)  & 0.917 (0.209) & 0.523 (0.165)  & 1.051 (1.130)  \\ 
             & 1024  & 0.600 (0.076)  & 0.930 (0.338) & 0.610 (0.244)  & 1.305 (1.325)   \\ 
             & 2048  & 0.652 (0.115)  & 1.062 (0.483) & 0.735 (0.326)  & 1.906 (2.389)   \\  \hline
             \multirow{4}{*}{\centering Beta$(1,1)$}
             & 128   & 3.473 (0.632)  & 1.467 (0.348)  & 0.757 (0.186)  & 0.969 (0.550)  \\ 
             & 512   & 0.887 (0.094)  & 1.041 (0.267)  & 0.582 (0.215)  & 1.348 (1.587) \\
             & 1024  & 0.641 (0.090)  & 1.029 (0.481)  & 0.686 (0.378)  & 1.573 (1.841) \\ 
             & 2048  & 0.669 (0.117)  & 1.227 (0.687)  & 0.865 (0.504)  & 2.462 (2.880) \\ \hline  
             \multirow{4}{*}{\centering Beta$(5,5)$}
             & 128   & \textbf{3.006} (0.599)  & \textbf{1.276} (0.286)  & 0.680 (0.169)  & 0.811 (0.422)  \\ 
             & 512   & 0.781 (0.084)  & 0.923 (0.229)  & 0.536 (0.188)  & 0.914 (0.778) \\ 
             & 1024  & 0.589 (0.086)  & 0.873 (0.327)  & 0.594 (0.247)  & 1.306 (1.473) \\ 
             & 2048  & 0.624 (0.098)  & 1.116 (0.815)  & 0.790 (0.565)  & 1.662 (2.096) \\ \hline                
\end{tabular}
\end{table}

\begin{figure}
    \centering
    \includegraphics[width=.9\linewidth]{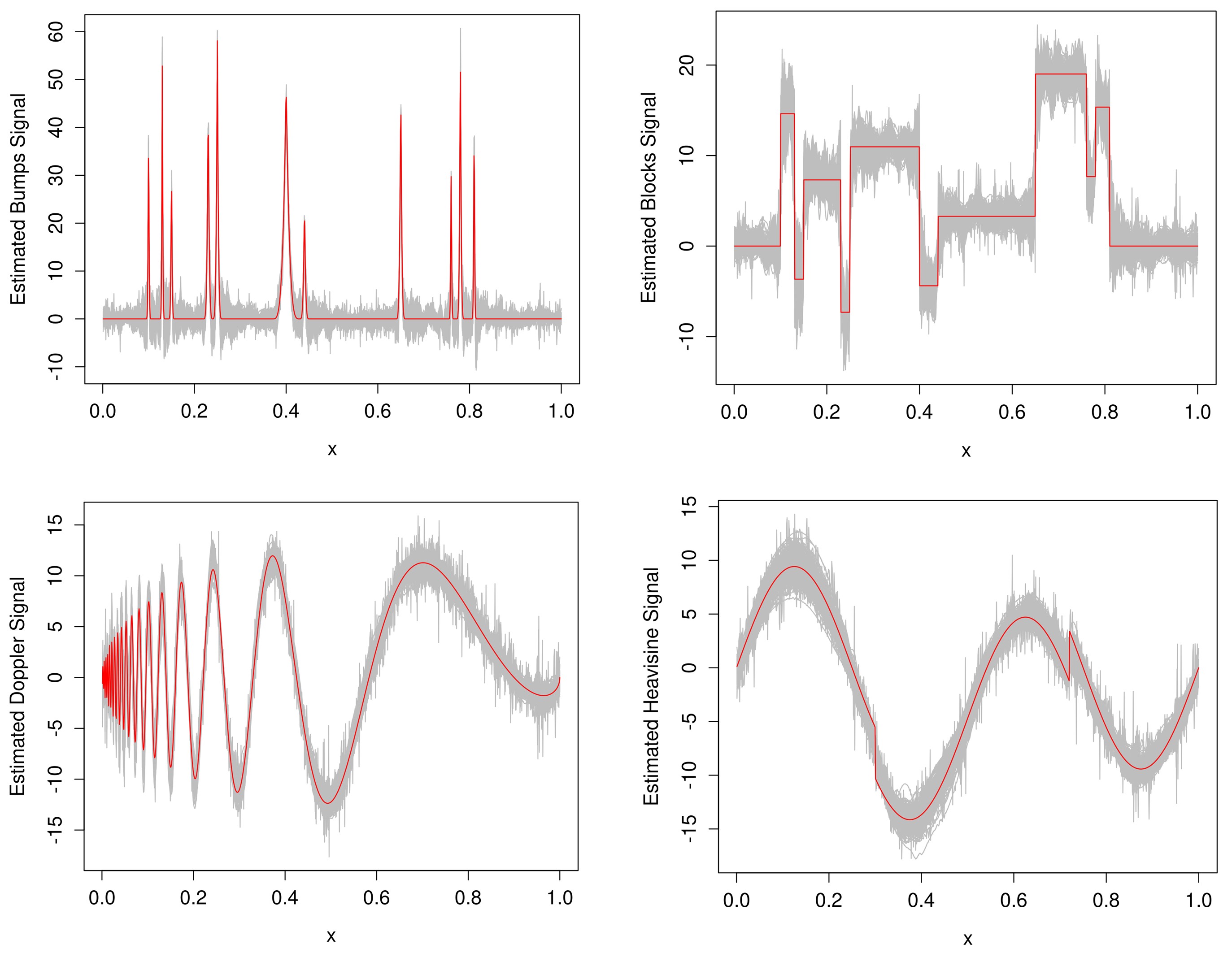} 
    \caption{$200$ replications (in gray) of the estimates obtained by the application of the shrinkage rule under raised cosine prior for $n=1024$ and $\mathrm{SNR} = 3$. The red curve is the underlying function.}
    \label{rep:app}
\end{figure}

Figure \ref{rep:app} displays, as an example, the estimated signals for $n = 1024$ and $\mathrm{SNR} = 3$ across the 200 replications conducted in the simulation study, compared to the actual signal (test function). The gray curves represent the obtained estimates, while the red curve represents the actual signal. It is evident that the proposed estimator is effective in capturing the primary characteristics of the functions, such as peaks and discontinuities, but faces challenges in regions where the behavior is more regular and smooth. In the case of the Heavisine function, for example, even though it has, by definition, smooth oscillations, the estimates obtained in each replication reveal more pronounced variability in peak and valley regions. This affects the values of the mean squared error and the median absolute error, impacting the performance of the estimators for each specific function.

\section{Real data applications}
Wavelet shrinkage has been broadly applied in fields that require noise removal and the preservation of relevant data features, particularly in signal processing and image analysis. Specifically, in the context of signal processing, there are several application areas, including time series treatment and neural activity signals. 

We illustrate the application of the shrinkage rule under the raised cosine prior in two real datasets. The first application deals with time series of temperature anomalies, while the second addresses noisy signals from electroencephalograms (EEG).

\subsection{Temperature anomalies dataset}

\begin{figure}
    \centering
    \includegraphics[width=1\linewidth]{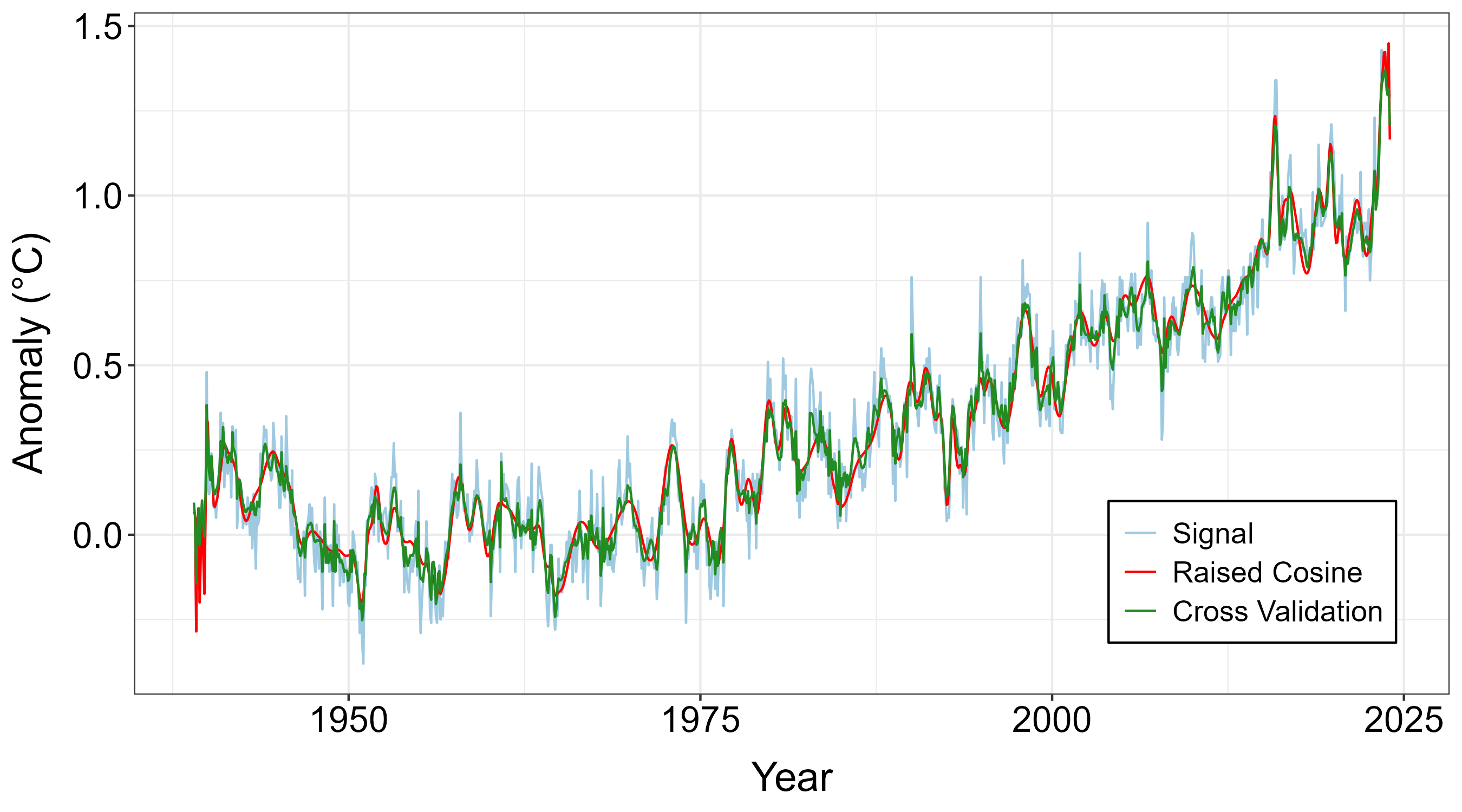} 
    \caption{Temperature anomalies dataset and the estimated signal using the raised cosine shrinkage rule (in red) and the cross-validation thresholding (in green).}
    \label{temp:app}
\end{figure}

 We considered the monthly time series of global temperature anomalies (in ºC) from January 1939 to April 2024, resulting in a sample size of $n = 1024$. The anomalies were calculated relative to the global average recorded between 1901 and 2000 and were obtained from the website of the United States National Oceanic and Atmospheric Administration (NOAA): \url{https://www.ncei.noaa.gov/access/monitoring/climate-at-a-glance/global/time-series/globe/tavg/land_ocean/1/0/1939-2024?filter=true&filterType=binomial}. The dataset is shown in Figure \ref{temp:app} (in blue).

\begin{figure}
    \centering
    \includegraphics[width=1\linewidth]{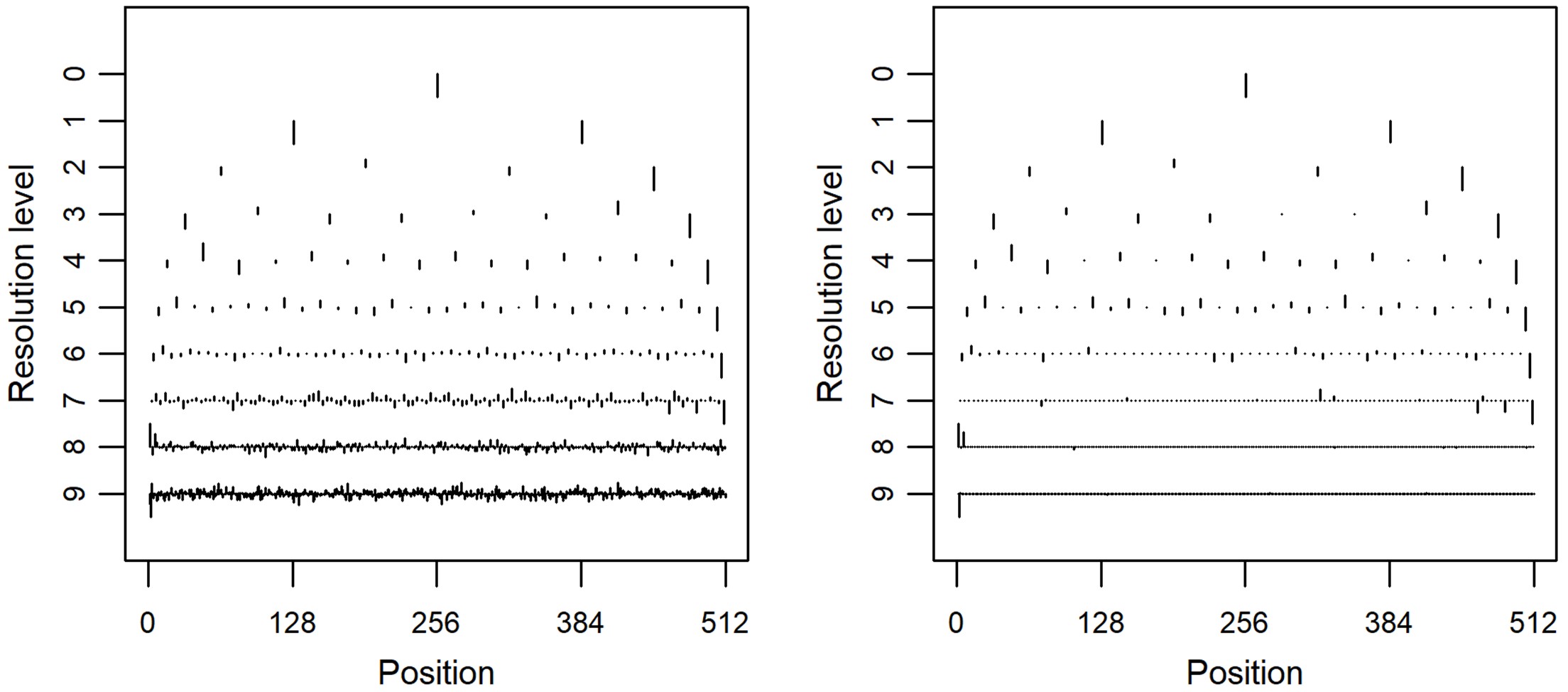} 
    \caption{Empirical (left panel) and estimated (right panel) wavelet coefficients by resolution level of the temperatute anomalies dataset.}
    \label{empcoef:app}
\end{figure}

Thus, the DWT was applied to the dataset using the Daub10 wavelet basis, resulting in the empirical wavelet coefficient vector. The shrinkage rule, based on the raised cosine prior, was applied to these coefficients to estimate the wavelet coefficients. We used $\alpha = 0.9$, $\hat{\sigma} = 0.07$ and $\tau = 2.79$, according to \eqref{eq:est-sigma} and \eqref{eq:tau}. Figure \ref{empcoef:app} illustrates the empirical coefficients (left panel) and the estimated (shrunk) wavelet coefficients (right panel) by resolution level, in which is possible to see the difference in the magnitude of the empirical coefficients caused by shrinking those predominantly composed of noise, mainly present at higher resolution levels, such as $j = 9$.

Subsequently, the IDWT was applied to the estimated coefficients, returning them to the original data domain and enabling signal reconstruction. The estimated signal using the raised cosine shrinkage rule (in red) is compared to the original dataset and the estimated signal using CV thresholding (in green) in Figure \ref{temp:app}. Furthermore, the estimated signal-to-noise ratio of the original series was $\widehat{\mathrm{SNR}} = 4.61$. This value reflects the behavior of the original signal, characterized by a few rapid variations and a moderate noise level. 

It is noted that the estimated series is capable of smoothing out fluctuations associated with finer scales in the time-scale decomposition, capturing general trends and patterns in temperature anomalies while attenuating higher-granularity details. In this context, noise removal better highlights the long-term structure while minimizing short-term fluctuations that are less relevant to the overall behavior of the series. 

\subsection{EEG dataset}

The EEG is an examination that records the electrical activity of the brain using electrodes placed on the scalp, monitors and detects brain wave patterns, helping in the diagnosis of conditions such as epilepsy. Video EEG, a variation of the test, combines the recording of electrical activity with a visual recording of the patient, allowing clinical events (such as seizures) to be correlated with changes in brain waves. This combination provides a more comprehensive analysis, which, in addition to enabling more accurate diagnoses, contributes to the evaluation and classification of epileptic seizures, guiding the most appropriate treatment strategies and evaluating the feasibility of possible surgery (see \cite{Sirven2016}). 

For this reason, in examinations aimed at classifying and characterizing epileptic seizures, in which the patient experiences convulsions, the signals related to the individual's neural activity might be contaminated by different types of artifacts, such as movement artifacts. These artifacts arise from interference with the sensitivity of the electrodes, which, under certain conditions, become more susceptible to noise. In this context, wavelet shrinkage methods are convenient for removing noise effects. 

In this context, we utilized a dataset comprising EEG recordings from 14 patients, collected at the Unit of Neurology and Neurophysiology of the University of Siena, Italy. Among the participants, there were nine men, aged between 25 and 71 years, and five women, aged between 20 and 58 years. The EEG recordings were obtained with a sampling rate of $512Hz$, meaning 512 observations per second. Additionally, during the examinations, various electrodes were positioned in different regions of the patients' heads, with each electrode identified by a code corresponding to the anatomical area in which it was located, forming the so-called EEG "channels". The data was obtained from \url{https://physionet.org/content/siena-scalp-eeg/1.0.0/#files-panel}.

Regarding the dataset used, examples of its application in recent research corroborate its reliability and widespread acceptance within the scientific community. In \cite{siena-scalp}, this dataset was used to propose a seizure prediction method based on the detection of synchronization patterns in EEG signals. Meanwhile, \cite{Dan-2024} used the same data to develop a validation framework for automated seizure detection algorithms. In another study, \cite{Marin-2024} explored the dataset to identify seizure onset zones in patients with drug-resistant epilepsy. 

\begin{figure}
    \centering
    \includegraphics[width=.8\linewidth]{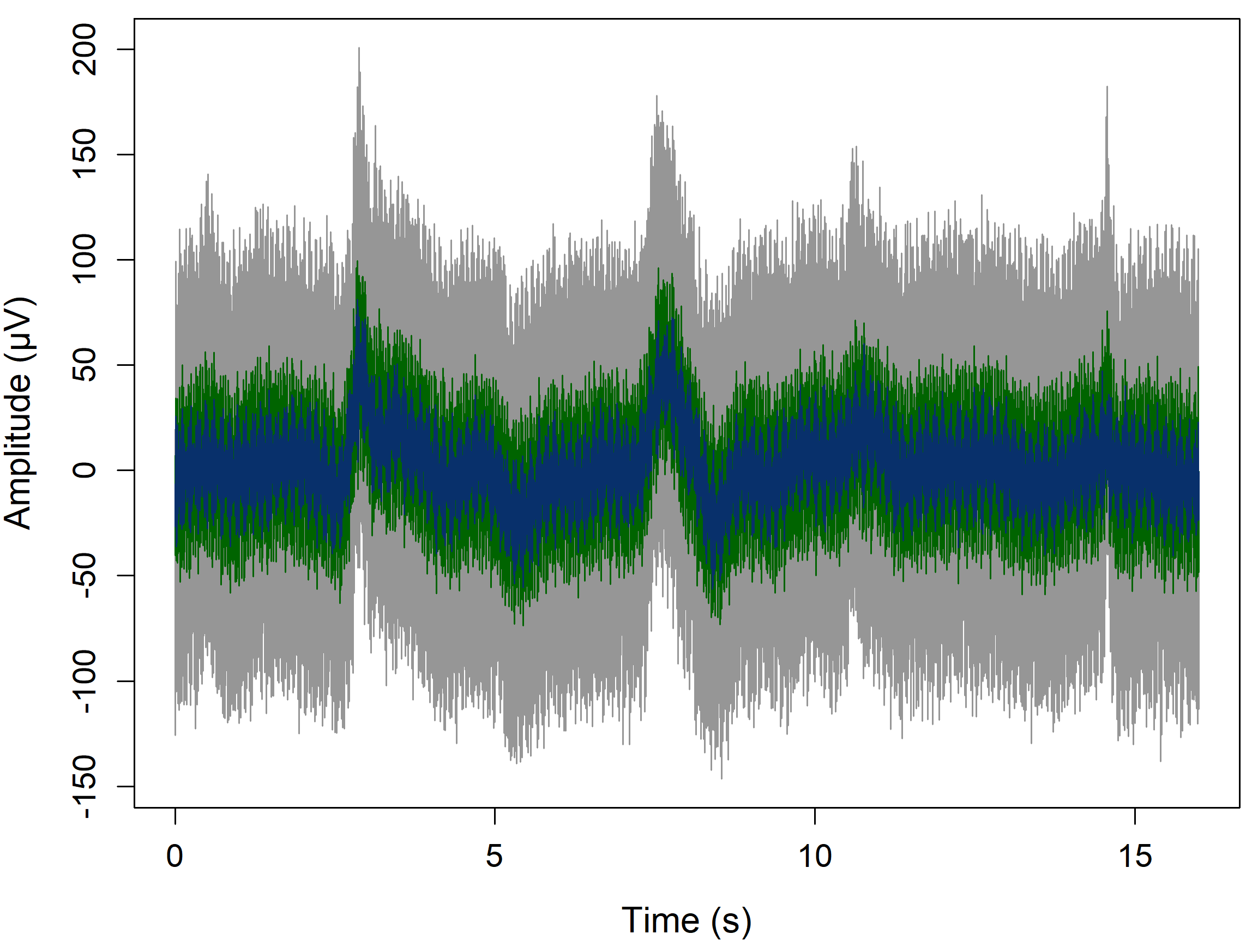} 
    \caption{Electroencephalogram (EEG) dataset (in gray) and the estimated signal using the raised cosine shrinkage rule (in blue) and the cross-validation thresholding (in green).}
    \label{eeg:app}
\end{figure}

In this context, the EEG signal corresponding to the recording of the third seizure of patient PN05, a 51-year-old woman, was selected. This signal is based on recordings from the Fp1 channel, which corresponds to the frontal polar region of the left cerebral cortex. It is also worth noting that, with a sampling rate of $512Hz$, obtaining the full extent of the signal would require a very large sample size. For this reason, a dyadic sample of size $n = 2^{13}$ was considered, covering $16$ seconds of the EEG recording, as illustrated in Figure \ref{eeg:app} (in gray).

Using the previously described procedure for shrinking empirical wavelet coefficients and maintaining the same estimation methods for the hyperparameters, the EEG signal was smoothed using the raised cosine prior shrinkage rule. In Figure \ref{eeg:app}, the smoothed signal is represented (in blue) and compared to the original dataset and the smoothed signal by CV thresholding (in green). Furthermore, the estimated hyperparameters and the signal-to-noise ratio were: $\widehat{\sigma} = 67.71$, $\tau = 530.28$, and $\widehat{\text{SNR}} = 0.37$. This result is consistent with the graphical analysis of the EEG signal, which exhibits high noise density, as evidenced by intense oscillations and rapid amplitude variations, making it difficult to distinguish the underlying signal without a smoothing process. 

The smoothed signal reveals a significant reduction in the extreme variability present in the original signal, attributed to rapid noise oscillations. Despite the predominance of noise in this case, which makes it challenging to recover detailed patterns or characteristics of the underlying signal, the application of the shrinkage rule proved effective in mitigating its impact. As the higher-magnitude coefficients preserve important local information, they highlight the structural aspects of brain activity over time.

\section{Final considerations}
In this study, we propose the use of a mixture of a point mass function at zero and a symmetric cosine distribution centered at zero as a prior distribution for the wavelet coefficients in a nonparametric regression problem. The associated Bayesian shrinkage rule was compared to other shrinkage rules developed under bounded support priors. Additionally, classical thresholding methods are considered for comparison with the proposed rule, to assess its performance under various scenarios defined in a simulation study.

The results obtained from both simulation experiments and real data applications demonstrate that the proposed shrinkage rule performs comparably to established methods in the literature, particularly in cases involving signals with low signal-to-noise ratios. Under such conditions, the raised cosine prior outperforms all classical approaches and competes directly with the rule based on the Beta prior. It may even outperform the latter in specific settings, such as with the Doppler function at SNR = 3, regardless of sample size. Further, its hyperparameters are directly related to the shrinkage level imposed by the rule on the empirical coefficients. 

These findings support the asymptotic behavior of the raised cosine distribution, which approximates the minimax rule over the class of symmetric and unimodal priors. Its superior performance in high Bayes risk settings or low signal-to-noise scenarios highlights the robustness of the proposed approach in practical applications.

\bibliographystyle{plainnat}
\bibliography{ref}

\clearpage

\clearpage

\clearpage

\appendix

\section{Proof of the Proposition 1}
Let is denote the likelihood function by $\mathcal{L} (.|\theta)$. Then, under the squared error loss function, the Bayes rule $\delta(\cdot)$ is the posterior mean of $\theta|d$, i.e.,
    \begin{align*}
        \delta(d) & = \mathbb{E}_{\pi} (\theta | d) = \int_{\Theta} \theta \pi(\theta|d) \, d\theta = \int_{\Theta} \theta \frac{\pi(\theta) \mathcal{L}(d|\theta)}{\int_{\Theta} \pi(\theta) \mathcal{L} (d|\theta) \, d\theta} \, d\theta \\[0.5cm]
        & = \frac{\int_{\Theta} \theta \left[ \alpha \delta_0(\theta) + (1-\alpha) g(\theta) \right] \mathcal{L}(d|\theta) \, d\theta}{\int_{\Theta} \left[ \alpha \delta_0(\theta) + (1-\alpha) g(\theta) \right] \mathcal{L}(d|\theta) \, d\theta} \\[0.5cm]
        & = \frac{\alpha \int_{\Theta} \theta \delta_0(\theta)\mathcal{L}(d|\theta) \, d\theta + (1-\alpha) \int_{\Theta} \theta g(\theta) \mathcal{L}(d|\theta) \, d\theta}{ \alpha \int_{\Theta} \delta_0(\theta) \mathcal{L}(d|\theta)  \, d\theta + (1-\alpha) \int_{\Theta} g(\theta) \mathcal{L}(d|\theta) \, d\theta} \\[0.5cm]
        & = \frac{(1-\alpha) \int_{\Theta} \theta g(\theta) \mathcal{L}(d|\theta) \, d\theta}{\alpha \mathcal{L}(d|\theta = 0) + (1-\alpha) \int_{\Theta} g(\theta) \mathcal{L}(d|\theta) \, d\theta} \\[0.5cm]
        & = \frac{(1-\alpha) \int_{-\tau}^{\tau} \theta g(\theta) \frac{1}{\sqrt{2 \pi} \sigma} \exp{\left\{ -\frac{1}{2} \left(\frac{d-\theta}{\sigma}\right)^2 \right\}} \, d\theta}{\alpha \frac{1}{\sqrt{2 \pi} \sigma} \exp{\left\{ -\frac{1}{2} \left(\frac{d}{\sigma}\right)^2 \right\}} + (1-\alpha) \int_{-\tau}^{\tau} g(\theta) \frac{1}{\sqrt{2 \pi} \sigma} \exp{\left\{ -\frac{1}{2} \left(\frac{d-\theta}{\sigma}\right)^2 \right\}} \, d\theta},
    \end{align*}

\vspace{0.3cm}
\noindent where in the last step we did $\mathcal{L} (.|\theta)$ the normal density function since $d|\theta \sim N(\theta, \sigma^2)$ and $g(\theta)$ the raised cosine density function \eqref{RC}. Doing $u = \frac{\theta - d}{\sigma}$, we have that 
\begin{align}
    \delta(d) & = \frac{(1-\alpha) \int_{\frac{-\tau-d}{\sigma}}^{\frac{\tau-d}{\sigma}} (\sigma u + d) g(\sigma u + d) \frac{1}{\sqrt{2 \pi} \sigma} \exp{\left\{ -\frac{1}{2} u^2 \right\}} \sigma \, du}{\alpha \frac{1}{\sqrt{2 \pi} \sigma} \exp{\left\{ -\frac{1}{2} \left(\frac{d}{\sigma}\right)^2 \right\}} + (1-\alpha) \int_{\frac{-\tau-d}{\sigma}}^{\frac{\tau-d}{\sigma}} g(\sigma u + d) \frac{1}{\sqrt{2 \pi} \sigma} \exp{\left\{ -\frac{1}{2} u^2 \right\}} \sigma \, du} \notag \\[0.5cm]
    & = \frac{(1-\alpha) \int_{\frac{-\tau-d}{\sigma}}^{\frac{\tau-d}{\sigma}} (\sigma u + d) g(\sigma u + d) \phi(u) \, du}{\frac{\alpha}{\sigma} \phi \left(\frac{d}{\sigma}\right) + (1-\alpha) \int_{\frac{-\tau-d}{\sigma}}^{\frac{\tau-d}{\sigma}} g(\sigma u + d) \phi(u) \, du} \notag \\[0.5cm]
    & = \frac{(1-\alpha) \int_{\frac{-\tau-d}{\sigma}}^{\frac{\tau-d}{\sigma}} (\sigma u + d) \frac{1}{2\tau} \left[1 + \cos\left(\frac{\pi (\sigma u + d)}{\tau}\right)\right]\phi(u) du}{\frac{\alpha}{\sigma} \phi(\frac{d}{\sigma}) + (1-\alpha) \int_{\frac{-\tau-d}{\sigma}}^{\frac{\tau-d}{\sigma}} \frac{1}{2\tau} \left[1 + \cos\left(\frac{\pi (\sigma u + d)}{\tau}\right)\right] \phi(u) du}. \label{eq:final}
\end{align}
Denoting the integrals of the numerator and denominator of \eqref{eq:final} by $I_{\text{num}}$ and $I_{\text{den}}$ respectively, we have that
\begin{align}
    I_{\text{num}} & = \int_{\frac{- \tau - d}{\sigma}}^{\frac{\tau - d}{\sigma}} (\sigma u + d) \frac{1}{2 \tau} \phi(u)  \, du 
    + \int_{\frac{- \tau - d}{\sigma}}^{\frac{\tau - d}{\sigma}} (\sigma u + d) 
    \frac{1}{2 \tau} \cos\left(\frac{\pi (\sigma u + d)}{\tau}\right) \phi(u) \, du \notag \\[0.5cm] 
    & = \frac{\sigma}{2 \tau} \int_{\frac{- \tau - d}{\sigma}}^{\frac{\tau - d}{\sigma}} u \, \phi(u) \, du 
    + \frac{d}{2 \tau} \int_{\frac{- \tau - d}{\sigma}}^{\frac{\tau - d}{\sigma}} \phi(u) \, du \notag = \int_{\frac{- \tau - d}{\sigma}}^{\frac{\tau - d}{\sigma}} (\sigma u + d) 
    \frac{1}{2 \tau} \cos\left(\frac{\pi (\sigma u + d)}{\tau}\right) \phi(u) \, du \notag \\[0.5cm]
    & = \frac{\sigma}{2 \tau} \left[ \phi \left( \frac{- \tau - d}{\sigma} \right) - \phi \left( \frac{\tau - d}{\sigma} \right) \right] 
    + \frac{d}{2 \tau} \left[ \Phi \left( \frac{\tau - d}{\sigma} \right) - \Phi \left( \frac{- \tau - d}{\sigma} \right) \right] \notag \\[0.5cm]
    & \quad + \int_{\frac{- \tau - d}{\sigma}}^{\frac{\tau - d}{\sigma}} (\sigma u + d) 
    \frac{1}{2 \tau} \cos\left(\frac{\pi (\sigma u + d)}{\tau}\right) \phi(u) \, du \notag \\[0.5cm]
    & = \frac{\sigma}{2 \tau} \left[ \phi \left( \frac{- \tau - d}{\sigma} \right) - \phi \left( \frac{\tau - d}{\sigma} \right) \right] 
    + \frac{d}{2 \tau} \left[ \Phi \left( \frac{\tau - d}{\sigma} \right) - \Phi \left( \frac{- \tau - d}{\sigma} \right) \right] + I_1, \label{eq:int-num}
\end{align}
where $I_1 =  \int_{\frac{- \tau - d}{\sigma}}^{\frac{\tau - d}{\sigma}} (\sigma u + d) 
    \frac{1}{2 \tau} \cos\left(\frac{\pi (\sigma u + d)}{\tau}\right) \phi(u) \, du$. Now, for $I_{\text{den}}$, it follows that
\begin{align}
    I_{\text{den}} & = \int_{\frac{- \tau - d}{\sigma}}^{\frac{\tau - d}{\sigma}} \frac{1}{2 \tau} \phi(u)  \, du + \int_{\frac{- \tau - d}{\sigma}}^{\frac{\tau - d}{\sigma}} \frac{1}{2 \tau} \cos\left(\frac{\pi (\sigma u + d)}{\tau}\right) \phi(u) \, du \notag \\[0.5cm] 
    & = \frac{1}{2 \tau} \left [\Phi \left( \frac{\tau - d}{\sigma} \right) - \Phi \left( \frac{- \tau - d}{\sigma} \right) \right] + \int_{\frac{- \tau - d}{\sigma}}^{\frac{\tau - d}{\sigma}} \frac{1}{2 \tau} \cos\left(\frac{\pi (\sigma u + d)}{\tau}\right) \phi(u) \, du \nonumber \\[0.5cm]
    & = \frac{1}{2 \tau} \left [\Phi \left( \frac{\tau - d}{\sigma} \right) - \Phi \left( \frac{- \tau - d}{\sigma} \right) \right] + I_2, \label{eq:int-den}
    \end{align}
where $I_2 = \int_{\frac{- \tau - d}{\sigma}}^{\frac{\tau - d}{\sigma}} \frac{1}{2 \tau} \cos\left(\frac{\pi (\sigma u + d)}{\tau}\right) \phi(u) \, du.$
\vspace{0.5cm}
Finally, substituting \eqref{eq:int-num} and \eqref{eq:int-den} in \eqref{eq:final}, we have that the proposed shrinkage rule under raised cosine prior is given by
\begin{equation}
      \delta(d) = \frac{(1-\alpha) \frac{\sigma}{2 \tau} \left [\phi \left( \frac{- \tau - d}{\sigma} \right) - \phi \left( \frac{\tau - d}{\sigma} \right) \right] + \frac{d}{2 \tau} \left [\Phi \left( \frac{\tau - d}{\sigma} \right) - \Phi \left( \frac{- \tau - d}{\sigma} \right) \right] + I_1} {\alpha \frac{1}{\sigma} \phi(\frac{d}{\sigma}) + (1-\alpha) \frac{1}{2 \tau} \left [\Phi \left( \frac{\tau - d}{\sigma} \right) - \Phi \left( \frac{- \tau - d}{\sigma} \right) \right] + I_2}. \nonumber  \qed
\end{equation}

\end{document}